\documentclass[aps,pra,twocolumn,showpacs,superscriptaddress,floatfix]{revtex4}  
\usepackage{graphicx}  
\usepackage{amssymb}  
\usepackage{epsfig}  
\usepackage{amsmath}  
\usepackage{dcolumn}
\usepackage{bm}
\usepackage{bbm}  
\usepackage{color}  
\usepackage{hyperref}  
\usepackage[up]{subfigure}

\newcommand{\be}{\begin{equation}}  
\newcommand{\ee}{\end{equation}}  
\newcommand{\bc}{\begin{center}}  
\newcommand{\ec}{\end{center}}  
\newcommand{\bea}{\begin{eqnarray}}  
\newcommand{\eea}{\end{eqnarray}}  
\newcommand{\ba}{\begin{array}}  
\newcommand{\ea}{\end{array}}

\begin{document}    
\title{Hamilton's turns as visual tool-kit   
for designing of single-qubit unitary gates}   
\author{B. Neethi Simon}   
\email{neethisimon@gmail.com}  
\affiliation{Department of Mechanical Engineering,   
SSN College of Engineering, OMR, SSN Nagar, Chennai 603 110}  
\author{C. M. Chandrashekar}   
\email{cmadaiah@phys.ucc.ie}  
\affiliation{Optics  \& Quantum Information Group, The Institute of   
Mathematical Sciences, Tharamani, Chennai 600 113} 
\affiliation{Ultacold Quantum Gases, Physics Department,
National University of Ireland, UCC, Cork, Ireland}   
\author{Sudhavathani Simon}   
\email{sudhasimon@gmail.com}  
\affiliation{Department of Computer Science,   
Women's Christian College,  Chennai 600 006, India}  
  
  
\begin{abstract}  
 Unitary evolutions of a qubit are traditionally represented geometrically   
as rotations of the Bloch sphere, but the composition of such evolutions   
is handled algebraically through matrix multiplication   
[of SU(2) or SO(3) matrices].    
Hamilton's construct, called turns, provides for handling the latter pictorially through the as addition of directed great circle arcs on the   
unit sphere S$^2 \subset \mathbb{R}^3$, resulting in a non-Abelian version of the  
parallelogram law of vector addition of the Euclidean translation group.  
This construct is developed into a visual   
tool-kit for   
handling the design of single-qubit unitary gates. As an application,   
 it is shown, in the concrete case wherein the qubit is realized as   
polarization states of light, that all unitary gates can be realized   
conveniently through   
 a universal gadget consisting of just two quarter-wave plates (QWP)    
and one half-wave plate (HWP). The analysis and results easily   
transcribe to other realizations of the qubit: The case of NMR   
is obtained by simply substituting $\pi/2$ and   
$\pi$ pulses respectively for QWPs and HWPs, the phases of the pulses  
playing the role of the orientation of fast axes of these plates.       
\end{abstract}  
\maketitle  

\section{Introduction}  
States of a qubit are in one-to-one correspondence with  points   
of the (solid) unit ball $D^3\subset \mathbb{R}^3\,$($D$ for disk);   
depending on the context it is  called the Poincar\'{e} or Bloch ball.  
Pure states are on the boundary  S$^2$ of $D^3$ and mixed states  
correspond to the interior points. The von Neumann entropy which is  
a measure of the mixedness of the state $\rho$ has the simple form  
\begin{eqnarray*} 
S(\rho) = -\sum_{\pm}\,\frac{1\pm r}{2} {\rm log}_2\frac{1\pm r}{2},\;\;0\le r\le 1,   
\end{eqnarray*} 
where $r$ is the radial distance of the point representing the state  
measured from the center of $D^3$. When the qubit is realized as  
polarization states of light, $r$ equals the degree of  
polarization\,\cite{oneil,wolf}. 
The center corresponds to the maximally mixed (completely unpolarized) state.  
\par 
To go with this attractive geometric portraying of states, unitary evolutions  
$\rho \to U\rho\,U^\dagger,~U\in $ SU(2) act as (three-dimensional) rotations  
leaving the center of the state space $D^3$ invariant. This is a realization of  
the adjoint representation of SU(2) as the two-to-one SU(2) $\to$ SO(3)  
homomorphism, both $U$ and $-\,U$ of SU(2) imaging to the same element  
of SO(3).  
 More general physical evolutions or channels act as {\em inhomogeneous  
linear maps} on $D^3$;  
in addition to mapping $D^3$ {\em into} itself, some additional requirement amounting  
to {\em complete positivity}\,\cite{davis-book} will have to be satisfied by these  
maps. Such nonunitary evolutions, however, play no role in this work.  
\par  
Though  
states and their (unitary) evolutions are thus represented {\em geometrically},  
composition or concatenation of evolutions is traditionally handled {\em  
algebraically} through matrix multiplication. Hamilton's  
turns\,\cite{hamilton-lectures} offer a {\em visual tool} for handling  
this last aspect too  
in a geometrical or vivid pictorial manner. In this picture, unitary evolutions are  
represented by (equivalence classes of) directed great circle arcs on S$^2$, with  
composition of unitary evolutions correctly represented by a {\em geometric addition  
rule} for these directed arcs, quite analogous to the manner in which translation  
group elements in an Euclidean space are composed using the {\em parallelogram law} of  
vector addition.   
\par  
An extensive description of Hamilton's construct may be found  
in the book of Biedenharn and Louck\,\cite{biedenharn-book}, while a simplified  
presentation of the addition rule for turns is given more recently in  
Ref.\,\cite{turns-sc}. An early application of this construct to polarization optics  
can be found in\,\cite{turns-pramana}, and generalization of the construct to other  
low-dimensional groups can be be found in\,\cite{turns-prl,turns-jmp,turns-juarez,turns-vssc}.   
\par  
The principal aim of the present work is to develop Hamilton's geometric construct  
into a tool-kit for handling the composition and synthesis of unitary single-qubit  
gates in an efficient pictorial manner with no recourse to matrix multiplication.  
To be concrete, it is assumed in much of the presentation that our qubit  
is realized as  polarization states of a  
photon\,\cite{photon-qubit1,photon-qubit2,kimble}, but transliteration  
to other realizations  
of the qubit will be evident. For instance, the case of NMR quantum computation is
  easily seen to correspond to quarter-wave plates (QWPs)  and half-wave plates (HWPs) being replaced, respectively, by $\pi/2$ and $\pi$ pulses, the phases of the pulses
  playing the role of the orientation of the fast axis of the plates.

\par 
The  
tool-kit presented has the obvious limitation that it can handle only single-qubit  
(unitary)  
gates. Even so it could prove to be of value to quantum computation in view of the  
fact that single-qubit gates, along with {\em just one} two-qubit gate like the C-NOT  
gate, can realize all multi-qubit  
gates\,\cite{nielsen,gates-deutsch,gates-barenco,gates-divincenzo}.   
\par  
The usefulness of this tool-kit is not limited to the domain of quantum information   
and computation. The group SU(2) pervades many areas of science, either directly or  
through the rotation group ${\rm SO(3)} = {\rm SU(2)}/Z_2$, and this tool-kit  
could therefore prove useful in these other areas as well. In particular, it is of  
direct interest to classical polarization optics.   
\par 
The entire  
 presentation shoots towards the main result formulated as a theorem at the end of  
the paper,  
which asserts that all single-qubit gates can be conveniently realized using a  
universal gadget consisting of just two QWPs and one HWP.  
We should hasten to add, however, that this theorem is not new in itself, but has been  
formulated earlier using {\em algebraic methods}\,\cite{gadget-minimal}, and Bagini {\em et  
al.}\,\cite{gadget-gori} have presented a particularly helpful exposition of this result  
of Ref.\,\cite{gadget-minimal} which has been variously  
used\,\cite{used1,used2,used3,used4,used5,used6,used7}.   
Whereas the algebraic approach took a sequence of several 
papers\,\cite{turns-pramana,gadget-bhandari,gadget-universal} to  
eventually arrive at the final result 
 in the fourth\,\cite{gadget-minimal}, through a sequesnce of  
false starts and  
refinements, the  
pictorial approach presented here will be seen to render the result visual, and almost  
obvious.  
\par
Since this  geometric construct of Hamilton, called {\em turns} does not  
appear to be as well known as it deserves to be, we begin with a description of this  
construct itself, relating it to the three prominent parametrizations of SU(2)---the  
homogeneous Euler, the axis-angle, and the Euler parametrizations---and bringing out  
its interesting connection with the Berry-Pancharatnam geometric phase. 
  
\section{ HAMILTON'S TURNS}  
 
Reversible gates acting on a qubit  
{\em are in one-to-one correspondence} with  
 $2\times 2$ unitary matrices $u\in$ SU(2).   
The SU(2) matrices can be conveniently described by any triplet   
$\tau_{1}, \tau_{2}, \tau_{3}$ of Pauli-like Hermitian matrices satisfying   
the {\em defining algebraic relations}   
\bea  
\label{eq:1} 
\tau_{k}\tau_{l}=   
\delta_{kl}\tau_{0} + i \epsilon_{klm} \tau_{m},   
\eea  
where $\tau_{0}=   
\mathbbm{1}_{2\times2}$ is the unit matrix $\in$ SU(2).   
To be specific,  we   
take these matrices to be $\tau_{1}=\sigma_{3}, \tau_{2}=\sigma_{1}$ and   
$\tau_{3}=\sigma_{2}$, where $\sigma_{j}$'s are the standard Pauli   
matrices. The family of all unitary (reversible single-qubit)   
gates $u\in$ SU(2) then get parametrized as   
\bea   
\label{u}  
u &=& a_{0}\tau_{0}-i \bm{a} \cdot \bm{\tau} = \left( \ba{clcr}  
                                  a_{0}-ia_{1} & & -ia_{2} -a_3 \\  
                                  \\  
                                  -ia_{2} +a_3 & & a_{0}+ia_{1}  
  
                                  \ea \right), \nonumber\\  
&&\nonumber\\  
&&a_0^2 + a_1^2 +a_2^2 +a_3^2 \equiv a_{0}^2 + \bm{a} \cdot \bm{a} =1.  
\eea   
That is, the four real parameters $(a_{0}, \bm{a})=(a_{0}, a_{1}, a_{2}, a_{3})$,  
called the {\em homogeneous Euler parameters}, correspond to a  point on the  
three-sphere S$^{3} \subset \mathbb{R}^{4}$. Elements of SU(2) are thus in one-to-one  
correspondence with points on S$^{3}$, consistent with and exhibiting the fact that    
S$^{3}$ is the group manifold of SU(2). Hamilton's turns constitute a powerful  
visual representation of this S$^{3} \subset {\mathbb R}^4$ 
  on S$^{2} \subset {\mathbb R}^3$ through (equivalence  
classes of) directed great circle arcs, with group multiplication of  
SU(2) matrices (concatenation of  single-qubit unitary gates) {\em faithfully  
transcribed} into a  "parallelogram law of addition" for these directed geodesic  
arcs on S$^{2}$.   
\par   
Given a unitary gate $u$ or $(a_{0}, \bm{a}) \in$ S$^3$, the constraint $a_{0}^2 + \bm{a} \cdot \bm{a} =1$ guarantees that we can find an ordered pair of unit   
vectors $\bm{\hat{n}}_1 ,\, \bm{\hat{n}}_2 \in \mathbb{R}^{3}$ such that   
\be   
\label{a}   
a_{0}= \bm{\hat{n}}_1 \cdot \bm{\hat{n}}_2, ~~~~~ \bm{a} = \bm{\hat{n}}_1\wedge   
\bm{\hat{n}}_2.   
\ee   
It follows that a directed great circle (or geodesic) arc on S$^{2}$, with tail at  
$\bm{\hat{n}}_1$ and head at $\bm{\hat{n}}_2$, can be associated with the unitary gate $u$. Clearly,   
such an association is {\em not unique}, and the non-uniqueness is   
precisely to the following extent: Any pair $\bm{\hat{n}}_1' ,\, \bm{\hat{n}}_2'$   
obtained by rotating {\em both} $\bm{\hat{n}}_1$ and $\bm{\hat{n}}_2$ by {\em equal} amount   
about $\bm{\hat{n}}_1 \wedge \bm{\hat{n}}_2$ will meet the requirements in Eq.\,(\ref{a}), and   
hence will correspond to the gate represented by the original pair   
$\bm{\hat{n}}_1, \bm{\hat{n}}_2$. Such a rotation obviously corresponds to   
 {\em rigidly sliding} the directed arc representing $u$ along its great   
circle.   
\par  
One is thus led to  consider {\em equivalence classes of directed   
great circle arcs} on S$^2$, the equivalence being with respect to   
the sliding just noted: Two such directed arcs are equivalent if they   
are on the same great circle and if one can be made to coincide with the other by rigidly   
sliding it on the great circle. These equivalence classes are called {\em Hamilton's   
turns}.  It is clear that  elements of SU(2) are in {\em one-to-one   
correspondence} with turns (assuming the arclength of turns is   
restricted not to exceed $\pi$).   
\par  
If $u\in$ SU(2) is represented by the turn whose representative element is the directed great   
circle arc from $\bm{\hat{n}}_1$(tail) to $\bm{\hat{n}}_2$(head), we    
denote this fact through   
\be \label{u1} u=T(\bm{\hat{n}}_1,   
\bm{\hat{n}}_2) = \bm{\hat{n}}_1 \cdot \bm{\hat{n}}_2 - i \bm{\hat{n}}_1 \wedge   
\bm{\hat{n}}_2 \cdot \bm{\tau}.   
\ee   
Henceforth we talk of turns, SU(2) matrices, and unitary single-qubit gates interchangeably. For brevity, we often call a representative arc itself as the turn and its length as the length of the turn or simply as the {\em turn length}, but this abuse   
of terminology should cause no confusion.   
\par  
Two special elements of SU(2) namely $\tau_{0}$ and $ -\tau_{0}$ are   
distinguished in that they constitute the {\em center} of the group. This   
distinction should be expected to manifest itself in any   
representation, and the one due to Hamilton happens to be no exception.   
The unit element $\tau_{0}$, the trivial gate, corresponds to the null turn   
$\bm{\hat{n}}_1 = \bm{\hat{n}}_2 \in$ S$^{2}$,   
 and the gate $\,-\tau_{0}$ to $\bm{\hat{n}}_1 = - \bm{\hat{n}}_2 \in   
S^{2}$, the respective turn-lengths being $0,\,\pi$. The equivalence   
class associated with either is clearly a {\em two-parameter family},  
since $\bm{\hat{n}}_1$ can be any point on  S$^{2}$.    
  However, the equivalence class of directed great circle arcs associated with any other turn is a {\em one-parameter family}. In particular, every turn or $u\in\,$SU(2), $u\ne \pm   
\tau_0$, has associated with it a   
{\em unique directed great circle} of S$^2$.   
\par  
Analogy with the Euclidean translation group (in two dimensions, for instance) wherein group elements are represented by  equivalence classes of `{\em free  vectors}' is obvious. There the abelian group composition takes the geometric form of {\em parallelogram law} of vector addition. It turns out that such a geometric or pictorial composition of group elements applies to the present non-Abelian case of SU(2) as well, with turns playing the role of (equivalence classes of)  free vectors; indeed, the power of Hamilton's turns can be traced   
to this fascinating fact.   
\begin{figure}[ht]  
\includegraphics[width=6.5cm]{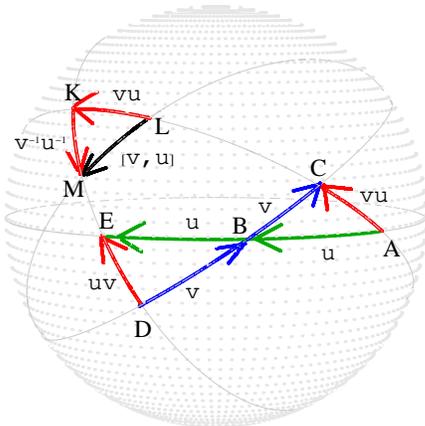}  
\caption{\footnotesize{The "parallelogram law" or   
"addition rule" for turns, with   
${\rm turn~AB\;=\;turn~BE}$ and   
${\rm turn~BC\;=\;turn~DB}$ representing, respectively, unitary gates   
$u$ and $v$ and ${\rm turn~AC}$ and ${\rm turn~DE}$ representing,   
respectively, the products  $vu$ and $uv$. Since     
${\rm turn~LK}$ represents $vu$ and    
${\rm turn~KM}$, the inverse of  
${\rm turn~DE}$, represents $v^{-1}u^{-1}$, and     
 ${\rm turn~LM}$ represents the commutator $[v,\,u]=v^{-1}u^{-1}vu$. }}  
\label{fig1}  
\end{figure}   
\par 
To see this pictorial composition law, assume that we are   
given two SU(2) gates $u, v$ and we wish to compute geometrically (visually)    
the matrix product $vu$. Referring to Fig.\,\ref{fig1}, let the   
directed great circle arc AB represent $u$ and let BC represent $v$.   
It is important to note that the representative arcs are so chosen    
that the head of the right factor $u$ and the tail of the left    
factor $v$ coincide at B$\,\sim\,\bm{\hat{n}}_2$; since great circles on S$^{2}$  
certainly intersect, {\em this can always be arranged for any given pair of turns}.  
Now draw the directed geodesic arc from the free tail A$\,\sim\,\bm{\hat{n}}_1$ to the   
free head C$\,\sim\,\bm{\hat{n}}_3$ to obtain a new turn represented by the directed arc AC. The important claim is that  turn~AC correctly represents the matrix product $vu$. This   
is readily verified through matrix multiplication:  
\bea \label{vu}    
vu&=&   
T(\bm{\hat{n}}_2,\bm{\hat{n}}_3)T(\bm{\hat{n}}_1,\bm{\hat{n}}_2)\nonumber \\  
    &=& (\bm{\hat{n}}_2 \cdot \bm{\hat{n}}_3 - i \bm{\hat{n}}_2 \wedge   
\bm{\hat{n}}_3 \cdot \bm{\tau}) (\bm{\hat{n}}_1 \cdot \bm{\hat{n}}_2 - i   
\bm{\hat{n}}_1 \wedge \bm{\hat{n}}_2 \cdot \bm{\tau})\nonumber \\   
&=& \bm{\hat{n}}_1 \cdot   
\bm{\hat{n}}_3 - i \bm{\hat{n}}_1 \wedge \bm{\hat{n}}_3 \cdot \bm{\tau} =   
T(\bm{\hat{n}}_1, \bm{\hat{n}}_3).   
\eea  
\noindent 
{\bf Remark}: The only property of the $\tau$-matrices used in this   
verification is $\tau_{k}\tau_{l} = \delta_{kl}\tau_{0} + i\epsilon_{klm}\tau_{m}$, and hence this result and its consequences are independent of which set of Pauli-like matrices was actually used.   
\par  
We forced ourselves to perform the kind of matrix multiplication in   
Eq.\,(\ref{vu}), just once, simply to demonstrate  that {\em the one-to-one   
correspondence between} SU(2) {\em gates and turns is indeed a group isomorphism}.  
The rest of   
this work, however,  will rest solely on the geometric or visual construct of turns,  
with almost no recourse to matrix multiplication. We note in passing that the above   
composition law immediately implies that the matrix inverse of the SU(2)   
gate or turn $T(\bm{\hat{n}}_1, \bm{\hat{n}}_2$) corresponds to $T(\bm{\hat{n}}_2, \bm{\hat{n}}_1$), the {\em reversed turn.}  
\par  
To compute or construct the `other' product $uv$, choose E and D on the great   
circles, respectively, through A, B and C, B such that AB\;=\;BE and BC\;=\;DB. Then   
turn~BE\;=\;turn~AB represents $u$ and     
turn~DB\;=\;turn~BC represents $v$. It is thus obvious in view of    
Eq.\,(\ref{vu}), that   
turn~DE represents the product $uv$, which is manifestly different from   
the product $vu$ (represented by turn~AC), giving a vivid pictorial   
depiction of the noncommutative nature of the "addition rule" for    
turns, consistent with the non-Abelian nature of SU(2) composition.  
 
\vskip 0.2cm 
\noindent 
{\bf Remark}:  
In spite of being quite different from one another,  
turn~AB and turn~DE share  one important common aspect.   
To see this, consider the spherical triangles ABC and EBD. The angle at   
 B is the same for both triangles, AB\;=\;BE, and CB\;=\;BD. The two   
triangles are thus congruent, showing that turn~AC and turn~DE have   
{\em the same turn length}.  As we shall see, this is a pictorial  
manifestation of the fact ${\rm tr}\,vu   
= {\rm tr}\,uv$.    
\par  
Presented also in Fig.\,\ref{fig1} is  a visual display of the {\em   
commutator } of two SU(2) gates. Recall that the commutator of a pair of elements  
$g_1,\,g_2$ of a multiplicative group $G$ is defined as $[g_2,\,g_1] \equiv   
g_2^{-1}g_1^{-1}g_2g_1$, the {\em multiplicative difference} of $   
g_1g_2$ and $g_2g_1$. Let the geodesic arcs DE and AC when extended   
meet at K. Choose points L, M on these extended arcs such that AC\;=\;LK and   
ED\;=\;KM. Since turn~EB corresponds to $u^{-1}$ and turn~BD   
to $v^{-1}$, turn~ED corresponds to the product $v^{-1}u^{-1} = (uv)^{-1}$.  
Referring now to the spherical triangle LKM,   
since turn~LK corresponds to $vu$ and turn~KM to $v^{-1}u^{-1}$    
we deduce, again in view of Eq.\,(\ref{vu}), that turn~LM corresponds to the product     
$v^{-1}u^{-1}vu$,  the commutator $[v,\,u]$ of interest.   
\par  
It is often convenient to rewrite the SU(2) composition rule of matrix  
multiplication   
\bea  
\label{t1}  
T(\bm{\hat{n}}_3, \bm{\hat{n}}_4)   
T(\bm{\hat{n}}_2, \bm{\hat{n}}_3) T(\bm{\hat{n}}_1, \bm{\hat{n}}_2) =   
T(\bm{\hat{n}}_1, \bm{\hat{n}}_4)   
\eea  
as the geometric (visual) rule of "addition" of turns    
\be   
\label{t2}  
\mbox{turn}~\bm{\hat{n}}_1\bm{\hat{n}}_2 + \mbox{turn}~\bm{\hat{n}}_2   
\bm{\hat{n}}_3 + \mbox{turn}~\bm{\hat{n}}_3 \bm{\hat{n}}_4 =   
\mbox{turn}~\bm{\hat{n}}_1\bm{\hat{n}}_4.   
\ee   
In transcribing from the "multiplication" mode of Eq.\,(\ref{t1}) to the "addition"  
mode of   
Eq.\,(\ref{t2}), however, it is important to remember   
that the individual terms in this "sum" in Eq.\,(\ref{t2}) {\em read from left to   
right} correspond to the factors in the SU(2) matrix product Eq.\,(\ref{t1}) {\em   
read from right to left}; the order is important, the "sum" being   
non-commutative. This geometric "addition rule" for turns, which is   
clearly reminiscent of the parallelogram law for the composition of   
elements of the (Abelian) Euclidean translation group, is associative   
and faithfully represents the non-Abelian or non-commutative group composition in SU(2).    
\par  
Our consideration of turns so far has been based on the homogeneous Euler parametrization of SU(2) Eq.\,(2). The group SU(2) can also be parametrized in the axis-angle form   
\bea  
\label{un}  
u = u(\bm{\hat{n}},\alpha) &\equiv& \exp(-i\frac{\alpha}{2}\bm{\hat{n}}\cdot   
\bm{\tau})\nonumber\\  
  &=& \cos\frac{\alpha}{2}\, \tau_0  
     -i \sin\frac{\alpha}{2}\,\bm{\hat{n}}\cdot \bm{\tau}; \nonumber\\  
u(\bm{\hat{n}},2\pi) &=& -\tau_0,\;\;\;  
u(\bm{\hat{n}},4\pi) = \tau_0,\;\;\nonumber\\  
u(\bm{\hat{n}},2\pi + \alpha) &=& u(-\bm{\hat{n}},2\pi-\alpha),    
\eea  
where $\bm{\hat{n}}$ is a unit vector $\in {\mathbb R}^3$. We may,   
 in view of the last line of Eq.\,(\ref{un}), restrict   
$\alpha$ to the range $0\le \alpha\le 2\pi$.    
 A  view of turns   
which corresponds to this parametrization proves more   
convenient for some purposes, and so we  describe it briefly.  
\par 
We have seen that every turn, other than the special   
turns $T(\bm{\hat{n}},\bm{\hat{n}})$ and $T(\bm{\hat{n}},-\bm{\hat{n}})$   
corresponding to elements in the {\em center} of SU(2), has associated with it a {\em unique directed} great circle; this great circle and an angle (length of the representative arc)   
fully specifies the turn. However, directed great circles and directed axes  
(or unit vectors $\bm{\hat{n}}\in S^2$) are in one-to-one correspondence:   
if the directed great circle   
is specified by the ordered pair $(\bm{\hat{n}}_1,\bm{\hat{n}}_2)$ of linearly  
independent unit vectors, the directed axis is specified by the unit vector $\bm{\hat{n}}\in S^2$ in the direction of $\bm{\hat{n}}_1\wedge \bm{\hat{n}}_2$. We may thus denote a turn  alternatively by the symbol $T(\bm{\hat{n}},\alpha)$, with the understanding that the representative   
directed arc is on the great circle orthogonal to $\bm{\hat{n}}$ and has   
arclength $\alpha$. In other words, $T(\bm{\hat{n}},\alpha)$  corresponds to $u(\bm{\hat{n}},2\alpha)\in\,$ SU(2). With the two special turns excluded, it is clear that   
this representation is unique [\,$\bm{\hat{n}}_1\wedge \bm{\hat{n}}_2$ is nonvanishing for every $u$ not in the center of SU(2), that is, for every turn whose turn-length$~\in (0,\pi)\,$].    
\par
 We note from the last line of Eq.\,(\ref{un}) that $T(\bm{\hat{n}},\pi + \alpha)   
= T(- \bm{\hat{n}},\pi - \alpha)$, and hence the restriction of $\alpha$, the turn-length, to the range $0\le \alpha \le \pi$. For the two special turns corresponding to the center of SU(2), we see that $T(\bm{\hat{n}},0)$ and  $T(\bm{\hat{n}},\pi)$ represent, respectively,   
$\tau_0$ and $-\tau_0$, {\em independent} of $\bm{\hat{n}}\in S^2$.    
\par 
We note in passing a direct relationship between the turn length and the trace of the associated SU(2) matrix. Indeed, the facts that ${\rm tr}\,u(\bm{\hat{n}},\alpha) = 2 \cos   
\frac{\alpha}{2}$ and $u(\bm{\hat{n}},2\alpha)\in\,$SU(2) is represented by the turn $T(\bm{\hat{n}},\alpha)$ of length $ \ell_u =\alpha$  show that ${\rm tr}\,u = 2\cos \ell_u$. In particular, ${\rm tr}\,u$ is positive or negative depending on whether the turn length $\ell_u$   
 is $<$ or $> \pi/2$ : All traceless SU(2) matrices correspond to   
turn-length $\pi/2$, the popular Hadamard gate\,\cite{nielsen} and HWPs considered below being examples.   
 
\section{Realization of qubit as polarization states of light}  
 
As indicated earlier, our illustrations demonstrating the power of turns in   
solving problems of synthesis of unitary gates will use the concrete context wherein   
qubit is realized as the polarization  states of light. However, it is evident   
from the treatment to follow that the entire analysis applies equally well to other    
realizations of qubits and SU(2) gates, like NMR quantum computation\,\cite{nielsen} or   
passive (lossless) linear optics of a pair of radiation modes at lossless  
beam splitters\,\cite{saleh}.   
\par 
Birefringent media play a particularly dominant role in polarization optics, both classical and quantum. Consider a (quasi-monochromatic) light beam propagating  along the positive $x_{3}$ direction of a Cartesian system $(x_{1}, x_{2}, x_{3})$. The components $E_{1}, E_{2}$ of the transverse electric field along the $x_{1}, x_{2}$ directions can be arranged into a column vector $E \in {\mathcal C}^2$, called the Jones vector of the polarization state   
of the beam\,\cite{oneil}. The intensity equals $|E_{1}|^2 + |E_{2}|^2 = E^{\dagger}E$. A linear optical system is correspondingly represented by a $2 \times 2$ numerical matrix $J$ called the Jones matrix, and the input-output relationship is represented by   
\bea  
E_{\rm in} \rightarrow E_{\rm out} = J E_{\rm in}.   
\eea  
Lossless linear systems conserve intensity:   
$E^{\dagger}_{\rm out}E_{\rm out} = E^{\dagger}_{\rm in} E_{\rm in}$. It   
follows   
that the Jones matrices of such systems are unitary. Birefringent   
media, which introduce a relative phase between a characteristic pair   
of orthogonal  linear polarization states, and optically active   
media which introduce a relative phase between the two (orthogonal)   
circular polarization states are examples of such lossless linear   
systems of interest to polarization optics. Suppressing an overall   
phase, the Jones matrices of lossless linear systems can be identified with elements of the   
unimodular unitary group SU(2). In the case wherein the qubit corresponds  
to the polarization states  of a photon, these are indeed the relevant  
unitary or reversible single qubit gates.   
\par 
A birefringent plate (compensator) whose "fast axis" is along the   
transverse $x_{1}$ direction has the Jones matrix   
\bea   
\label{j}  
J=C_{0}(\eta)   
&=& \left( \ba{clcr}  
                              e^{-i\eta/2} & 0 \\  
                                0 & e^{i\eta/2}  
                       \ea \right)\nonumber\\   
&=& \exp\left (-i   
\frac{\eta}{2} \tau_{1}\right ) \in {\rm SU(2)},   
 \eea   
$\eta$ being the relative phase  introduced by the plate; we have $\eta =   
\epsilon\,\ell/\lambda$, where $\ell$ is the thickness of the plate,   
$\epsilon>0$ is the difference between the refractive indices for the   
two  characteristic orthogonal linear polarizations, and $\lambda$ is   
the wavelength. It is clear  that if the fast axis is at an angle $\varphi$ with   
the $x_{1}$ axis, then the Jones matrix would  be   
\bea   
J &=& C_{\varphi}(\eta) =  
\Phi(\varphi)C_{0}(\eta)\Phi(\varphi)^{-1} \nonumber\\  
 &=& \cos(\eta/2)\,\tau_0 \,-i\,\sin(\eta/2)  
\,(\cos(2\varphi)\,\tau_1 + \sin(2\varphi)\,\tau_2),\nonumber\\   
\eea  
where the two-dimensional matrix   
\be \Phi(\varphi)=\exp(-i\varphi\tau_{3})=\left(   
\ba{clcr}  
                                                      \cos\varphi &   
&-\sin\varphi \\  
                                                      \sin\varphi &   
& \cos\varphi \ea \right)   
\ee  
is an element of the subgroup SO(2) $\subset$ SU(2).   
   
\begin{figure}[h]  
\includegraphics[width=7.85cm]{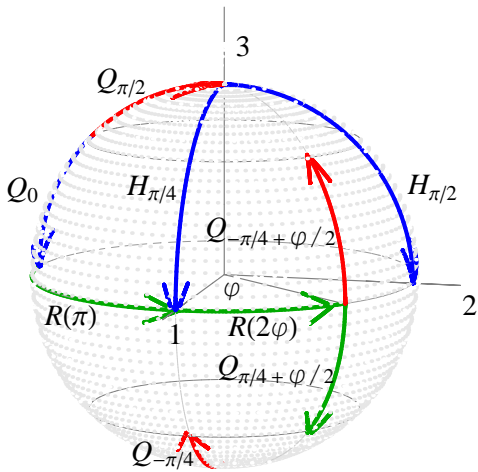}  
\caption{\footnotesize{The sphere of turns ${\cal T}$  
as it applies to the polarization qubit.  All  
turns on the equator (horizontal turns)  
correspond to optical rotators, and vertical turns correspond to birefringent plates.   
 While relating to the subscripts of HWPs and QWPs it should be   
remembered   
that spatial rotation of magnitude $\varphi/2$ is reflected on ${\cal T}$    
as polar rotation of magnitude $\varphi$ and not $\varphi/2$. }}  
 \label{fig2}   
\end{figure}   
Quarter-wave plates (QWPs) and half-wave plates (HWPs) are particular cases   
of birefringent plates and  correspond, respectively, to $\eta = \pi/2$ and $\pi$; they could equally well  be called $\lambda/4$ and $\lambda/2$ plates. The Jones matrix of a QWP   
with fast axis along the $x_{1}$-direction is thus $\exp(-i \frac{\pi}{4}   
\tau_{1}) = (\tau_{0}-i\tau_{1})/\sqrt{2}$; we denote this QWP by   
$Q_{0}$, so that $Q_{\varphi} = \Phi(\varphi) Q_{0} \Phi(\varphi)^{-1}    
 =[\,\tau_0 - i(\tau_1\cos 2\varphi +\tau_2 \sin 2\varphi)\,]/\sqrt{2}$ represents the QWP whose fast axis makes an angle $\varphi$ with the positive $x_{1}$ axis. Similarly, we use the notation $H_{0}$ for the HWP $-i\tau_{1}$ so that $H_{\varphi}$   
stands for $\Phi(\varphi) H_{0} \Phi(\varphi)^{-1} = - i(\tau_1\cos 2\varphi +\tau_2  
\sin 2\varphi)$, a HWP   
whose fast axis makes angle $\varphi$ with the $x_1$ axis. In particular, {\em $H_{\varphi/8}$ corresponds to the Hadamard gate}\,\cite{nielsen}. Thus $H_\varphi = (\,Q_\varphi\,)^2$, for all $\varphi$. These are particular cases of a more general and evident fact: If an    
optical system represented by Jones matrix $J$ is physically rotated by an angle $\varphi$ about the positive $x_{3}$ axis, the resulting system will have  Jones matrix $\exp(-i \varphi \tau_{3})J \exp(i\varphi \tau_{3})$.   
\par  
The SO(2) matrix $\exp(-i \varphi \tau_{3})$ plays yet another role in polarization optics: It is also the Jones matrix of an optically active medium. Specifically, an   
optically active medium (or simply optical rotator) which introduces a relative   
phase $\alpha$ between the left and the right circularly polarized   
states has the Jones matrix   
\bea \label{ux5}  
R(\alpha) = \exp(-i \frac{\alpha}{2} \tau_{3}).   
\eea  
Numerically, $R(\alpha) = \Phi(\alpha/2)$; however, we have chosen to   
use a different symbol $R(\alpha)$ for the optical rotator to distinguish   
it from $\Phi(\cdot)$, which stands for physical rotation of a gadget in the transverse plane.  
\par 
We depict in Fig.\,\ref{fig2} the sphere of turns ${\cal T}$.   
It is clear that all "vertical turns" correspond to birefringent plates:   
QWPs and HWPs have turn lengths of $\pi/4$ and $\pi/2$, respectively.   
Turns on the equator correspond to optical rotators.    
\par  
In addition to the homogeneous Euler  and axis-angle parametrizations   
of SU(2) gates already considered there exists a third one, the {\em Euler angle   
parametrization,}    
\bea  
 \label{u3}   
 u&=& u(\xi, \eta, \zeta) \nonumber \\ &\equiv& \exp(-i   
\frac{1}{2}\xi \tau_{3})\,\exp(-i \frac{1}{2}\eta \tau_{1})\,\exp(-i   
\frac{1}{2}\zeta \tau_{3}). ~~ ~~  
\eea   
This parametrization of the SU(2) group of unitary gates can be viewed as saying that every such gate is equivalent to an appropriate birefringent plate $C_0(\eta)$   
sandwiched between two appropriate optically active media $R(\xi),\,R(\zeta)$, with the effective thicknesses of the three media engineered to match the Euler   
parameters $\xi, \eta, \zeta$ of the gate under consideration.   
However, from an experimenter's point of view this cannot be the most   
convenient realization of the various unitary gates $J\in$\, SU(2); unlike the HWP and QWP, $C_0(\eta)$ is not a component readily available off the shelf, and  a variable $R(\xi)$  tends to introduce  $\xi$-dependent loss and dynamical phase. {\em It turns out that} QWPs {\em and} HWPs {\it alone are sufficient}.  As we will see, the geometric representation of Hamilton renders this fact particularly transparent and visual.  
\par  
We may note in passing that the Euler angle parametrization [Eq.\,(\ref{u3})] can be rewritten in the modified form   
\bea  \label{uxy}   
u(\xi,\eta,\zeta) = C_{\xi/2}(\eta)\,R({\xi+\zeta}) =   
R({\xi+\zeta})\,C_{-\zeta/2}(\eta).~~~  
\eea  
This means an arbitrary unitary gate is a variable birefringent plate   
preceded or followed by a variable rotator. The fact still remains that, unlike QWPs and HWPs, variable birefringent plates are nonstandard polarization optical components, and optically active media introduce undesirable losses and dynamical phases which vary with  the optical rotation or effective thickness of the medium.   
   
\section{Hamilton's turns and transformation of polarization states}   
\label{Hturns}   
  
Polarization states are conveniently described (even in the classical case) by restricting   
attention to Jones vectors of unit norm (unit intensity) and ignoring an overall phase. Such  normalized Jones vectors correspond, in the quantum case, to  state vectors of a {\em two-level system or qubit}. In either case, the  state  represented by a normalized Jones vector {\em E} is fully determined by the (complex) ratio $z = E_2/E_1$. This is rendered particularly transparent by going over to the coherency or density matrix, and one  then finds that   
polarization states are in one-to-one correspondence with points on the unit sphere  
S$^2$, called the Poincar\'e or Riemann sphere, obtained by identifying the points $z\to \infty$ (the one-point compactification) of the complex $z$ plane:   
\bea  
 \mbox{tr}(EE^{\dagger}) &=&   
1~~\mbox{and}~~ e^{i\alpha} E   
(\bm{\hat{m}}) \sim E(\bm{\hat{m}}),~~~\bm{\hat{m}} \in S^2, \nonumber \\    
&\rightarrow& \rho(\bm{\hat{m}})\equiv E(\bm{\hat{m}})E(\bm{\hat{m}})^{\dagger}.\;\, \nonumber  
\eea  
Written in more detail,   
\bea   
E(\bm{\hat{m}}) = \frac{1}{\sqrt{2}}   
\left(\ba{cr}  
              e^{-i \varphi/2}\cos \frac{\theta}{2}  + e^{ i \varphi/2} \sin   
\frac{\theta}{2}  \\  
              \\  
   \;i\left( e^{-i \varphi/2}\cos   
\frac{\theta}{2}  -  e^{i \varphi/2}\sin \frac{\theta}{2}\right)\;\\      
\ea\right),~~~  
\eea   
so that for any state $\bm{\hat{m}}\in{\rm S}^2$ we have for the ratio  
$z(\bm{\hat{m}})$ of the components of   
$E(\bm{\hat{m}})$   
the expression   
\be   
z(\bm{\hat{m}}) = \frac{E_{2}}{E_{1}} = \frac{\sin   
\theta \sin \varphi + i \cos \theta}{1 + \sin {\theta} \cos   
{\varphi}}.   
\ee   
The corresponding coherency or density matrix reads    
\bea  
\rho(\bm{\hat{m}}) &=& E(\bm{\hat{m}})E(\bm{\hat{m}})^\dagger \nonumber \\   
 &&\nonumber\\   
&=& \frac{1}{2} \left(\ba{clcr}  
  1+ \sin \theta \cos \varphi & & \;\sin \theta \sin \varphi - i   
\cos \theta\; \\  
              \\  
                \sin \theta \sin \varphi + i \cos \theta & & 1- \sin   
\theta \cos \varphi\\  
              \ea \right)\nonumber\\   
         &&\nonumber     \\  
 &=& \frac{1}{2}(\tau_{0} + \bm{\hat{m}} \cdot \bm{\hat\tau});\nonumber\\  
z(\bm{\hat{m}}) &=& \rho_{12}(\bm{\hat{m}})/\rho_{11}(\bm{\hat{m}}).   
\eea   
The parameters   
$\theta, \varphi$ are respectively the polar and azimuthal   
coordinates of $\bm{\hat{m}} =(\sin \theta \cos \varphi,\,  
\sin\theta \sin\varphi,\,\cos\theta) \in S^2$.   
  
Had we used in place of the "$\tau$ matrices" the standard Pauli   
$\sigma$ matrices,  the coherency matrix would have read   
\bea  
\rho(\bm{\hat{m}}) = \frac{1}{2}(\sigma_0 +   
 \bm{\hat{m}} \cdot \bm{\sigma}) = \frac{1}{2} \left( \ba{clcr}  
       1+ \cos \theta & & \;e^{-i \varphi}\sin \theta\;  \\  
              \\  
                e^{i \varphi}\sin \theta  & & 1- \cos \theta \\  
              \ea \right).\nonumber\\  
\eea  
[$\sigma_0$ again is the unit matrix $\mathbbm{1}_{2\times2}=\tau_0$.]   
Consequently,   
$E(\bm{\hat{m}})$ would have been parametrized as   
\bea   
E(\bm{\hat{m}}) = \left   
( \ba{clcr}   
\cos \theta/2 \\   
\\   
e^{i \varphi} \sin \theta/2 \ea \right ),   
\eea   
so that the ratio  $E_{2}/E_{1}$ becomes    
 $z(\bm{\hat{m}})= e^{i \varphi}\,\tan (\theta/2)$, a form more   
familiar in the context of NMR and the associated Bloch sphere.    
  
\begin{figure}[ht]  
\includegraphics[width=7.75cm]{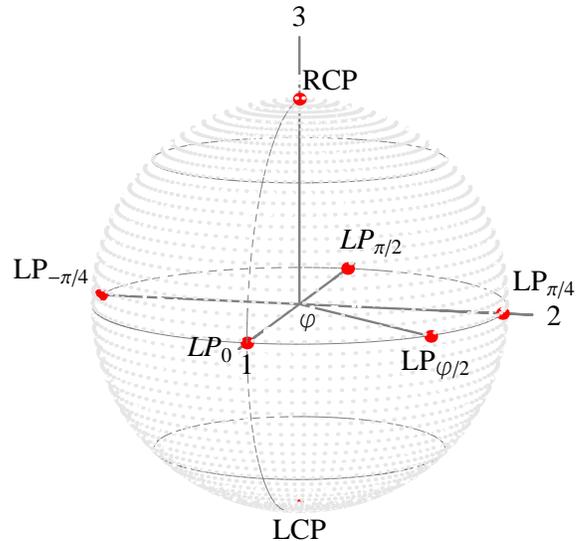}  
\caption{\footnotesize{The polarization states of light represented as points on the   
Poincar\'{e} sphere ${\cal P}$. Right and left circularly polarized states occupy   
the polar positions, and points on the equator correspond to linear polarization   
states. Spatial rotation of the plane of polarization by angle $\varphi$   
corresponds to $2\varphi$ rotation about the polar axis. Thus linear polarization   
 ${\rm LP}_0$ along the $x_1$ axis and linear polarization ${\rm LP}_{\pi/4}$   
 at $45^{\circ} $ to the $x_1$ axis are separated on ${\cal P}$ by   
$90^{\circ}$. Points of the upper (lower) hemisphere correspond,  
respectively, to right (left) handed elliptic polarization. }}  
 \label{fig3} \end{figure}   
\par 
Staying with the choice $\tau$, rather than $\sigma$, we sketch in   
Fig.\,\ref{fig3}  the Poincar\'e sphere $\mathcal P$. We adopt the   
convention that right and left circular polarization (RCP and LCP)   
states correspond, respectively, to $\frac{1}{\sqrt{2}}\left( \ba{clcr} 1 \\ i   
\ea \right)$ and $\frac{1}{\sqrt{2}}\left( \ba{clcr} 1 \\ -i   
\ea \right)$ or, equivalently, to the Poincar\'e sphere   
coordinates $(0, 0, 1)$ for RCP and $(0,0, -1)$ for LCP. Linear   
polarization at  angle $\alpha$ to the (positive) $x_{1}$ axis has   
Jones vector $\left( \ba{clcr} \cos \alpha \\ \sin \alpha   
\ea \right)$ and corresponds to the Poincar\'e sphere   
coordinates $(\cos 2\alpha, \sin 2\alpha, 0)$. Thus, all linear   
polarization states are on the equator; all states above the   
equator have right-handed elliptic polarization, while those in the   
lower hemisphere are left-handed. Antipodal points of $\mathcal P$   
correspond to orthogonal polarization states:   
 $E(-\bm{\hat{m}})^\dagger E(\bm{\hat{m}}) = 0$, equivalently   
${\rm tr}\,(\,\rho(-\bm{\hat{m}})\rho(\bm{\hat{m}})\,)=0$.   
 
\vskip 0.2cm 
\noindent 
{\bf Remark on convention:} Since the expressions for $E(\bm{\hat{m}})$ and $\rho(\bm{\hat{m}})$   
 may "appear to be" considerably simpler with the choice $\bm{\sigma}$, one may   
wonder why one chose $\bm{\tau}$ in the first place. The reason is one of  
convention, a price one occasionally pays for tradition. In the   
case of a spin-$\frac{1}{2}$ particle of the NMR context or a   
two-level atom, one   
chooses the vectors   
 $\left( \ba{clcr} 1 \\ 0 \ea \right)$   
and $\left( \ba{clcr} 0 \\ 1 \ea \right)$   
 to correspond to the poles of S$^2$ and prefers   
to associate Bloch's name with this sphere. Obviously, these   
 so-called computational   
basis states are eigenstates of $\sigma_{3}$. In polarization   
optics, on the other hand, the circularly polarized states   
$\frac{1}{\sqrt{2}}\left( \ba{clcr} 1 \\ \pm i \ea \right)$   
are given the special   
honor of polar positions, but these are eigenstates of   
$\sigma_{2} =\tau_3$.   
Following tradition we wish to keep the polar axis as the vertical   
{\em and} third axis for polarization qubit. Cyclic permutation of the $\sigma$ matrices seems to be the minimal way of meeting these concerns or requirements   
without offending in any way the basic algebra [\,commutation and   
anti-commutation relations, Eq.\,(\ref{eq:1})\,].

\section{Action of turns on the Poincar\'e sphere and connection with   
geometric phase}  
  
The notation $T(\bm{\hat{n}},\alpha)$ for turns, corresponding to the   
axis-angle parametrization   $u (\bm{\hat{n}},\,2\alpha) =   
\exp(-i{\alpha}\,\bm{\hat{n}}\cdot \bm{\tau})$ of SU(2),   
proves convenient in exhibiting the action of turns on the Poincar\'e   
sphere, our state space. Unitary gates $T(\bm{\hat{n}},\alpha)$ act on the Poincar\'e sphere,   
that is, on state $\rho(\bm{\hat{m}}) = \frac{1}{2}[\tau_0 + \bm{\hat{m}}\cdot \bm{\tau}]$   
represented by $\bm{\hat{m}}\in {\cal P}$, in this manner:  
\bea  
T(\bm{\hat{n}},\alpha):\;\;    
\rho(\bm{\hat{m}})   
&\to&  \rho(\bm{\hat{m}}^{\,\prime})=   
T(\bm{\hat{n}},\alpha)\, \rho(\bm{\hat{m}})\,  T(\bm{\hat{n}},\alpha)^{\,-1}\nonumber\\   
 &=& \exp(-i\,{\alpha}\,\bm{\hat{n}}\cdot \bm{\tau})   
\,\rho(\bm{\hat{m}})\, \exp(i\,\alpha\,\bm{\hat{n}}\cdot   
\bm{\tau}).\nonumber\\  
\eea  
As expected, the two special gates $\tau_0,\,-\tau_0$ corresponding to   
elements of the center of SU(2) are seen to have no effect on the Poincar\'e   
sphere and, as noted earlier, the set of parameters $(\bm{\hat{n}},\alpha)$   
in $T(\bm{\hat{n}},\alpha)$, with $0 <\alpha< \pi$, is unique for every   
other turn.    
\par 
Now, in view of the algebraic properties  of the   
$\tau$ matrices [Eq.\,(\ref{eq:1})], the above  transformation law simply reduces to      
\bea  
T(\bm{\hat{n}},\alpha): \; \;\bm{\hat{m}} \to    
\bm{\hat{m}}^{\,\prime} = (\bm{\hat{m}}\cdot\bm{\hat{n}})\bm{\hat{n}}  
 &+&   
 \cos 2\alpha [\bm{\hat{m}} - (\bm{\hat{m}}\cdot   
\bm{\hat{n}})\bm{\hat{n}}]\nonumber \\  
&+&\sin 2\alpha \,\bm{\hat{n}}\wedge\bm{\hat{m}}.  
\eea  
Note that $(\bm{\hat{m}}\cdot\bm{\hat{n}})\bm{\hat{n}}$   
 is the component of $\bm{\hat{m}}$ along $\bm{\hat{n}}$  
and $\bm{\hat{m}} - (\bm{\hat{m}}\cdot \bm{\hat{n}})\bm{\hat{n}}$  
is the component of  $\bm{\hat{m}}$ orthogonal to $\bm{\hat{n}}$   
and hence in the plane spanned by $\bm{\hat{m}}, \bm{\hat{n}}$.   
Further $\bm{\hat{n}}\wedge\bm{\hat{m}}$ which is orthogonal to   
both $\bm{\hat{m}}$ and $\bm{\hat{n}}$ has the same magnitude as  
$\bm{\hat{m}} - (\bm{\hat{m}}\cdot \bm{\hat{n}})\bm{\hat{n}}$.   
Thus, {\em the effect of $T(\bm{\hat{n}},\alpha)$ on the state space or Poincar\'{e} sphere ${\cal P}$ is an} SO(3) {\em rotation about $\bm{\hat{n}}$,     
of extent $2\alpha= $  {\em twice} the length of the turn}.   
In particular, the three one-parameter subgroups of SU(2) gates  
$\exp(\,-i\,\frac{\alpha_k}{2}\,\tau_k\,),\;\,k=1,2,3~$,   
 act on ${\cal P}$,  respectively, through the following   
one-parameter subgroups of SO(3) rotations:  
\bea   
\exp(\,-i\,\frac{\alpha_1}{2}\,\tau_1\,) \to    
{\cal R}_1(\alpha_1) = \left(\ba{ccc}  
1&0&0\\  
~0~&\cos\alpha_1&-\sin\alpha_1\\  
0&\sin\alpha_1&\cos\alpha_1  
\ea \right) \,,\nonumber\\  
\nonumber\\  
\exp(\,-i\,\frac{\alpha_2}{2}\,\tau_2\,) \to    
{\cal R}_2(\alpha_2) = \left( \ba{ccc}  
\cos\alpha_2&~0~&\sin\alpha_2\\  
0&1&0\\  
-\sin\alpha_2&0&\cos\alpha_2  
\ea \right) \,,\nonumber\\  
\nonumber\\  
\exp(\,-i\,\frac{\alpha_3}{2}\,\tau_3\,) \to    
{\cal R}_3(\alpha_3) = \left( \ba{ccc}  
\cos\alpha_3&-\sin\alpha_3&~0~\\  
\sin\alpha_3&\cos\alpha_3&0\\  
0&0&1  
\ea \right) \,.\nonumber\\  
\eea  
\par  
\begin{figure}[ht]  
\includegraphics[width=8.0cm]{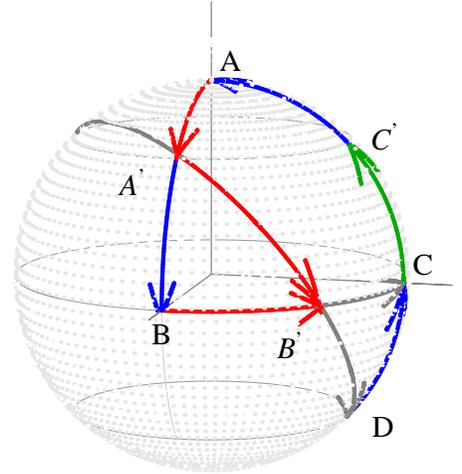}  
\caption{\footnotesize{Shown on the sphere of turns ${\cal T}$ is a geodesic triangle   
corresponding to a unitary resolution of identity. While ${\rm   
turns~AB,\;\;BC,\;\,CA}$ compose to unity, their respective square roots compose   
to ${\rm turn~B'C}$ whose turn length is half the area of the triangle. }}  
\label{fig4}  
\end{figure}   
\par  
Now consider the spherical triangle ABC on the sphere of turns ${\cal   
T}$ shown in Fig.\,\ref{fig4}, the coordinates of A,B,C being respectively   
$(0,0,1),\,(1,0,0),\,(0,1,0)$. Since   
turn AB corresponds to the unitary gate $-i\tau_2$,   
turn BC to $-i\tau_3$, and turn CA to $-i\tau_1$, the (closed) triangular circuit,    
\bea  
{\rm turn~AB} + {\rm turn~BC} +  
{\rm turn~CA} = {\rm null~turn}  
\eea  
represents a visual depiction of the matrix product identity   
\bea\label{ux}  
 (-i\tau_1)\,(-i\tau_3)\,( -i\tau_2) = \tau_0   
\eea  
involving the Pauli gates---, a unitary resolution of the identity.    
\par 
Let A$'$, B$'$, and C$'$ be the midpoints of AB, BC, and CA respectively.   
Thus,  ${\rm turn~AA'} = {\rm turn~A'B}$,    
${\rm turn~BB'}= {\rm turn~B'C}$ and ${\rm turn~CC'} = {\rm turn~C'A}$    
 are the square roots of   
 ${\rm turn~AB}$, ${\rm turn~BC}$, and ${\rm turn~CA}$, respectively.     
It is instructive to compute the composition of these square-root turns:   
${\rm turn~AA'} +  {\rm turn~BB'} + {\rm turn~CC'}$. Note that    
${\rm AA'}$, ${\rm BB'}$, and ${\rm CC'}$    
have   
equal arclength of $\pi/4$, and correspond respectively to  
$\frac{1}{\sqrt{2}}(\tau_0-i\tau_2)$,   
$\frac{1}{\sqrt{2}}(\tau_0-i\tau_3)$, and  
$\frac{1}{\sqrt{2}}(\tau_0-i\tau_1)$.   
Since ${\rm turn~AA'} = {\rm turn~A'B}$, we have ${\rm turn~AA'} +   
{\rm turn~BB'} = {\rm turn~A'B'}$.  
To compose ${\rm turn~A'B'}$ with ${\rm turn~CC'}$ extend ${\rm A'B'}$   
to meet the great circle   
through A and C at D. Comparing the spherical triangles   
${\rm BB'A'}$ and ${\rm CB'D}$, we see that ${\rm BB'} = {\rm B'C}$,   
${\rm angle~A'BB'} = {\rm angle~B'CD}$ $(=\pi/2)$, and   
${\rm angle~BB'A'} = {\rm angle~CB'D}$.   
The two triangles are thus congruent, and so ${\rm turn~A'B'} =   
{\rm turn~B'D}$ and arclength of ${\rm DC}$ = arclength of ${\rm A'B}$   
$(=\pi/4)$.  Thus, ${\rm turn~A'B'} + {\rm turn~CC'} = {\rm turn~B'D} + {\rm turn~DC} =   
{\rm turn~B'C}= {\rm turn~BB'}$. We have thus shown   
\bea \label{uy}  
{\rm turn~AA'} + {\rm turn~BB'} +  
{\rm turn~CC'} &=& {\rm turn~BB'},\nonumber\\  
{\rm i.\,e.\,,}\;\, \sqrt{-i\tau_1}\, \sqrt{-i\tau_3}\,   
\sqrt{-i\tau_2} = \sqrt{-i\tau_3}&=&   
\frac{1}{\sqrt{2}}(\tau_0-i\tau_3).\;\;\nonumber\\   
\eea  
The square roots thus  compose to produce neither the null turn nor its   
square root (any turn of turn length  =$\pi$), but   
$\frac{1}{\sqrt{2}}(\tau_0-i\tau_3)$,  a {\em primitive eighth root} of the   
null or identity turn $\tau_0$.   
\par 
This is due to the following fact: Unlike triangles in the Euclidean plane,   
there is no notion of {\em similar triangles} (beyond congruence)   
in the  spherical case; the area of a spherical triangle is   
fully determined by its three angles.   
This situation described by Eqs.\,(\ref{ux}) and (\ref{uy}) should   
be contrasted with the corresponding situation in   
respect of the Euclidean parallelogram law, wherein if three elements   
of the translation group compose to produce the null (identity), element   
then their respective "square roots" too will certainly compose to   
yield the null   
element, corresponding to a {\em similar triangle} with one-fourth   
area and the same interior angles as the original one. The failure of   
the SU(2) turns in this respect, as depicted   
by Eq.\,(\ref{uy}), is rooted in the non-Abelian nature of the group on the one hand  
and in the nontrivial curvature of S$^2$ (as compared to the Euclidean  plane) on the other; indeed,   
these two aspects go hand in hand, as may be seen also by consideration   
of the geometric or Pancharatnam phase.     
\par 
Before we turn to the geometric phase, we note, however, that the sum of the three   
(interior) angles of the triangle ABC is in excess of $\pi$ by $\pi/2$, and this   
{\em spherical excess} equals the area of the triangle. We note also   
that the turn length of the composite ${\rm turn~AA'} + {\rm turn~BB'} +  
{\rm turn~CC'}$, which is clearly a measure of the extent to which the   
square roots fail to compose to the null turn,  is $\pi/4$, precisely   
half of the area of the original (closed) triangle ABC. That is, in the  
axis-angle notation $T({\bm{\hat n}},\alpha)$, the value of $\alpha$ corresponding   
to composition of the square roots (that is, ${\rm turn~B'C}$) is $\pi/4$; and  
$\bm{\hat{n}}$  
corresponds   
precisely to $\bm{\hat{n}}_A$,  the "starting point" A of the triangle.   
\par  
All these aspects apply not only to the particular triangle shown in   
Fig.\,\ref{fig4} but also to an {\em arbitrary geodesic triangle};    
construction of  proof in the  general case is slightly more elaborate,  
but   
very similar to the one for the special triangle in Fig.\,\ref{fig4}\;(see Ref.\,\cite{two-level}). That is, we have for any spherical triangle ABC with area ${\cal A}$   
\bea  
{\rm turn~AA'} &+& {\rm turn~BB'} + {\rm turn~CC'} =   
 T(\bm{\hat{n}}_A,\,{\cal A}/2), \;\;\;\;  
\eea  
where ${\rm A',\;B',\; C'}$ are, respectively, the midpoints of AB, BC, CA and   
$\bm{\hat{n}}_A$ is the unit vector pointing in the direction of A\;$\in S^2$.   
\begin{figure}[h]   
\includegraphics[width=6.5cm]{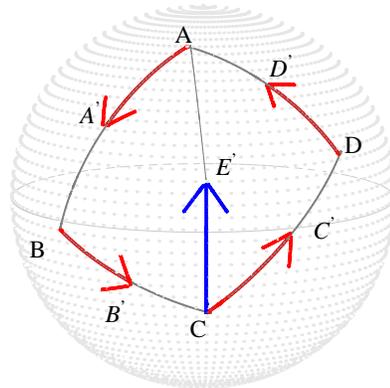}  
\caption{\footnotesize{Generalization of the area formula   
of Fig.\,\ref{fig4} to the case of a geodesic quadrilateral. }}  
\label{fig5}  
\end{figure}   
\par  
A simple argument  implies that {\em this area formula} applies indeed to all geodesic polygons. Consider, for instance, the (geodesic) quadrilateral ABCD on ${\cal T}$ shown in Fig.\,\ref{fig5}. Let ${\rm A'}$, ${\rm B'}$, ${\rm C'}$, ${\rm D'}$ be the mid points of   
AB, BC, CD, and DA, respectively, and let ${\rm E'}$ be the midpoint of   
the geodesic CA. Let ${\cal A}_1$ be the area of the triangle ABC and     
${\cal A}_2$  that of ACD so that  ${\cal A}= {\cal A}_1+{\cal A}_2$   
is the area of the quadrilateral, and let $\bm{\hat{n}}_A$ be the unit vector corresponding to the point A$\,\in {\cal P}$.  We have   
 the closed-circuit relation    
${\rm turn~AB} +{\rm turn~BC} +{\rm turn~CD} +{\rm turn~DA} = {\rm   
null~turn}$.   
We wish to compose the square roots of these turns in that order.    
The basic idea is to break this quadrilateral into two   
triangles ABC, ACD (making use of the fact that  
${\rm turn~CE'} +{\rm turn~AE'} \,=\, {\rm null~turn}$). We have   
\bea\label{uxx}  
{\rm turn~AA'} &+& {\rm turn~BB'} + {\rm turn~CC'} +{\rm   
turn~DD'}\nonumber\\   
 &=&({\rm turn~AA'} +{\rm turn~BB'} +{\rm turn~CE'})\nonumber\\   
 &&~~+ ({\rm turn~AE'} +{\rm turn~CC'} +{\rm turn~DD'}) \nonumber\\  
&=& T(\bm{\hat{n}}_A,\,{\cal A}_1/2) + T(\bm{\hat{n}}_A,\,{\cal A}_2/2)\nonumber\\  
&=& T(\bm{\hat{n}}_A,\,{\cal A}/2).   
\eea  
In the last step we used the fact that all turns $T(\bm{\hat{n}},\,\alpha)$   
 having the same $\bm{\hat{n}}$ form an (Abelian) one-parameter subgroup:   
\bea  
\label{uxxa} 
T(\bm{\hat{n}},\,\alpha)T(\bm{\hat{n}},\,\alpha^{\,\prime})  
= T(\bm{\hat{n}},\,\alpha^{\,\prime})T(\bm{\hat{n}},\,\alpha)  
 = T(\bm{\hat{n}},\,\alpha+ \alpha^{\,\prime}).\nonumber\\  
\eea  
Generalization to $n$-sided geodesic polygons is obvious.  
 
\vskip 0.2cm 
\noindent 
{\bf Remark}:  
As a subtle but important aspect we may note that the   
{\em addition rule} in Eq.\,(\ref{uxxa}) and its covariance under SU(2)   
conjugation (which rigidly translocates the quadrilateral   
on the sphere S$^2$) imply in turn the area formula   
in Eq.\,(\ref{uxx}), namely that the second argument of $T$ should   
necessarily be proportional to the area of the triangle ABC.   
\par  
Now we turn to the connection between geometric phase for two-level systems and the   
considerations of turns presented above. While geometric phase became   
popular owing to the seminal work of Berry\,\cite{berry1}, it had been   
"anticipated" by Pancharatnam\,\cite{pancha},  as pointed out by   
Nityananda and Ramaseshan\,\cite{nitya} and subsequently by   
Berry\,\cite{berry2}. Some of the other works which could be viewed, in retrospect,   
to have anticipated the geometric phase are listed in   
Ref. \,\cite{berry1}.   
\par 
In the course of his interference experiments with polarized light    
Pancharatnam faced this question: Given two distinct (nonorthogonal)   
polarization states represented by linearly independent Jones vectors   
$E(\bm{\hat{n}}_1)$ and $E(\bm{\hat{n}}_2)$, when should one say that these Jones vectors are in phase? Motivated by his experiments, Pancharatnam arrived at the following answer:      
 $E(\bm{\hat{n}}_1)$ and $E(\bm{\hat{n}}_2)$ are in phase if and only if the inner product   
$E(\bm{\hat{n}}_1)^\dagger E(\bm{\hat{n}}_2)$ is real positive.  
In other words,  {\em being in phase is synonymous with maximal constructive interference}.    
He noted that "being in phase" defined in this manner is {\em not an equivalence    
relation}, for it fails the transitivity requirement: $E(\bm{\hat{n}}_1)$ being in phase with $E(\bm{\hat{n}}_2)$ and $E(\bm{\hat{n}}_2)$ being in phase with $E(\bm{\hat{n}}_3)$   
does not necessarily imply that  $E(\bm{\hat{n}}_3)$ is in phase with $E(\bm{\hat{n}}_1)$.    
Indeed, he showed that this failure in respect of transitivity is geometric in nature, in the sense that if $E(\bm{\hat{n}}_1)$ is in phase with $E(\bm{\hat{n}}_2)$ and    
$E(\bm{\hat{n}}_2)$ is in phase with  $E(\bm{\hat{n}}_3)$, then $E(\bm{\hat{n}}_3)$  
will be necessarily out of phase with $E(\bm{\hat{n}}_1)$ precisely by half the area  
of the spherical triangle defined by vertices  
$\bm{\hat{n}}_1,\, \bm{\hat{n}}_2,\,\bm{\hat{n}}_3 \in {\cal P}$.    
\par 
As a simple illustration of this failure of transitivity, consider on the Poincar\'{e} sphere three points P, Q, R  identified by unit vectors $\bm{\hat{n}}_P = (0,\,1,\,0)$,    
$\bm{\hat{n}}_Q = (0,\,0,\,1)$, and $\bm{\hat{n}}_R = (1,\,0,\,0)$ corresponding, respectively, to linear polarization at an angle $\pi/4$ to the $x_1$ axis, RCP, and linear polarization along the $x_1$ axis. The  corresponding three Jones vectors may be taken to be    
\bea  
E(\bm{\hat{n}}_P)=\frac{1}{\sqrt{2}}\left( \ba{clcr} 1 \\1 \ea \right),&~\;~&  
E(\bm{\hat{n}}_Q) = \frac{1}{\sqrt{2}}e^{-i\,\pi/4}\left( \ba{clcr} 1 \\i \ea \right),\nonumber\\ E(\bm{\hat{n}}_R)&=&e^{-i\,\pi/4}\left( \ba{clcr} 1 \\0 \ea \right).  
\eea  
We have chosen the phase factors multiplying $E(\bm{\hat{n}}_Q)$,   
$E(\bm{\hat{n}}_R)$ so as to ensure  that    
$E(\bm{\hat{n}}_P)$ is in phase with $E(\bm{\hat{n}}_Q)$ and      
$E(\bm{\hat{n}}_Q)$  in phase with  $E(\bm{\hat{n}}_R)$.     
 It is readily verified that   
$E(\bm{\hat{n}}_R)$ is indeed out of phase with $E(\bm{\hat{n}}_P)$   
by $\pi/4$, half the area of the spherical triangle PQR.   
\par  
Given any three points $\bm{\hat{n}}_1$,  $\bm{\hat{n}}_2$,  $\bm{\hat{n}}_3$   
on the Poincar\'{e} sphere, we may write out this failure of transitivity in the transparent form   
\bea  
&&{\rm arg}\,[\,E(\bm{\hat{n}}_1)^\dagger E(\bm{\hat{n}}_2)\,      
E(\bm{\hat{n}}_2)^\dagger E(\bm{\hat{n}}_3)\,      
E(\bm{\hat{n}}_3)^\dagger E(\bm{\hat{n}}_1)\,] \nonumber\\  
&&={\rm arg}\,[\,{\rm tr}\,(\,E(\bm{\hat{n}}_1)E(\bm{\hat{n}}_1)^\dagger\,   
E(\bm{\hat{n}}_2)E(\bm{\hat{n}}_2)^\dagger\,   
E(\bm{\hat{n}}_3)E(\bm{\hat{n}}_3)^\dagger \,)\,]\;\, \nonumber\\  
&&= {\rm arg}\,[\,{\rm tr}\,(\,\rho(\bm{\hat{n}}_1)\,\rho(\bm{\hat{n}}_2)  
   \,\rho(\bm{\hat{n}}_3\,)\,]\nonumber\\      
&&~~\equiv \frac{1}{2}\,\Delta_3(\bm{\hat{n}}_1,\, \bm{\hat{n}}_2,\,\bm{\hat{n}}_3).      
\eea  
That $\Delta_3(\bm{\hat{n}}_1,\, \bm{\hat{n}}_2,\,\bm{\hat{n}}_3)$   
defined as above equals the area of the  spherical triangle with vertices at   
$\bm{\hat{n}}_1,\,\bm{\hat{n}}_2,\,\bm{\hat{n}}_3 \in {\cal P}$ is a result due to Pancharatnam. It may be noted that, $\Delta_3(\bm{\hat{n}}_1,\, \bm{\hat{n}}_2,\,\bm{\hat{n}}_3)$, the argument of the product of  pairwise inner products    
of "successive" states, is manifestly {\em gauge invariant} in the sense that   
if the Jones vectors $E(\bm{\hat{n}}_j)$ are replaced with $e^{i\alpha_j}E(\bm{\hat{n}}_j)$, where  $\alpha_{j},\;j=1,2,3$ are arbitrary phases, $\Delta_3(\bm{\hat{n}}_1,\,\bm{\hat{n}}_2,\,\bm{\hat{n}}_3)$ remains unaffected [\,every vector enters the expression as a bra and as a   
ket, once each\,].  It is in view of this gauge invariance that $\Delta_3$ may be called   
a geometric phase.   
\par  
If we are given four states  corresponding to points   
$\bm{\hat{n}}_1,\,\bm{\hat{n}}_2,\,\bm{\hat{n}}_3,\,\bm{\hat{n}}_4 \in {\cal P}$, the associated gauge-invariant geometric object is   
\bea \label{33}  
&&{\rm arg}\,[E(\bm{\hat{n}}_1)^\dagger E(\bm{\hat{n}}_2)      
E(\bm{\hat{n}}_2)^\dagger E(\bm{\hat{n}}_3)  
E(\bm{\hat{n}}_3)^\dagger E(\bm{\hat{n}}_4)E(\bm{\hat{n}}_4)^\dagger E(\bm{\hat{n}}_1)]\;\,\nonumber\\  
&&~~~~={\rm arg} [E(\bm{\hat{n}}_1)^\dagger E(\bm{\hat{n}}_2)      
E(\bm{\hat{n}}_2)^\dagger E(\bm{\hat{n}}_3)      
E(\bm{\hat{n}}_3)^\dagger E(\bm{\hat{n}}_1)]\nonumber\\  
&&~~~~~~~+\,   
{\rm arg}[E(\bm{\hat{n}}_1)^\dagger  E(\bm{\hat{n}}_3)      
E(\bm{\hat{n}}_3)^\dagger E(\bm{\hat{n}}_4)  
E(\bm{\hat{n}}_4)^\dagger E(\bm{\hat{n}}_1)]\nonumber\\  
&&~~~~\equiv \frac{1}{2}\Delta_4 (\bm{\hat{n}}_1,\,   
\bm{\hat{n}}_2,\,  
\bm{\hat{n}}_3,\,\bm{\hat{n}}_4),  
\eea  
where in the first step  we simply inserted into the expression    
$E(\bm{\hat{n}}_3)^\dagger E(\bm{\hat{n}}_1)\,  
E(\bm{\hat{n}}_1)^\dagger  E(\bm{\hat{n}}_3) =     
|\,E(\bm{\hat{n}}_3)^\dagger E(\bm{\hat{n}}_1)\,|^2 >0$, which does not affect the phase.   
That   
$\Delta_4(\bm{\hat{n}}_1,\, \bm{\hat{n}}_2,\,\bm{\hat{n}}_3,\,\bm{\hat{n}}_4)$  
has to be (a multiple of) the area of the quadrilateral follows also from the   
{\em additivity}    
\bea  
\Delta_4 (\bm{\hat{n}}_1,\bm{\hat{n}}_2,\bm{\hat{n}}_3,\bm{\hat{n}}_4)=   
 \Delta_3 (\bm{\hat{n}}_1, \bm{\hat{n}}_2,\bm{\hat{n}}_3) +  
 \Delta_3 (\bm{\hat{n}}_1, \bm{\hat{n}}_3,\bm{\hat{n}}_4)~~  
\eea  
demonstrated in Eq.\,(\ref{33}) and {\em independent of Pancharatnam}.   
That our considerations generalize to   
$n$ states and the associated $n$-sided polygon is clear.  
\par   
In the course of his celebrated proof of the Wigner theorem on   
symmetry in quantum theory Bargmann\,\cite{Bargmann} used the    
gauge-invariant expression $\Delta_3$ to discriminate between   
unitary and antiunitary symmetries;  and it is in honour of Bargmann that   
the authors of Ref.\,\cite{kinematic1,kinematic2}  named these     
invariants the  {\em Bargmann invariants}. Indeed, these authors showed that a very general theory of geometric phases can be formulated entirely on the basis of   
the Bargmann invariants\,\cite{kinematic1,kinematic2}. The Gouy phase, the phase jump a focused light beam suffers at the focal point, turns out to be a   
Bargmann invariant\,\cite{Gouy}, and this could probably be the   
earliest instance of a geometric phase observed in a laboratory.  
The connection between geometric phase and Bargmann invariants has been  
further  
explored in Refs.\,\cite{Barg-geo1,Barg-geo2,Barg-geo3,Barg-geo4}.  
\par  
With this preparation we are now ready to bring out the relationship between   
Hamilton's turns and the geometric phase through Bargmann invariants.   
We begin with two elementary observations. While "being in phase" is not an equivalence relation in general, on any geodesic arc of length $< \pi$ on ${\cal P}$   
it can indeed masquerade as one. For proof it suffices to note that the assertion is true for the particular case of Jones vectors of the form $\left(\,\begin{array}{c} \cos (\varphi/2)\\   
\sin (\varphi/2)\end{array}\,\right),\; 0\le\varphi <\pi$. On the   
Poincar\'{e} sphere this family occupies on the equator all points  with azimuthal   
coordinate $\varphi$ varying from $0$ up to (but not including) $\pi$.   
The fact that all these vectors (which correspond to linearly polarized states)   
are in phase with one another is obvious. That the claim in respect  
of masquerading applies to an arbitrary geodesic  
(of extent $<\pi$) on ${\cal P}$ follows  from the fact that  
all such geodesics on ${\cal P}$ are unitarily equivalent and  
from the fact that   
the very notion of being in phase is unitarily invariant,  
 since it is defined through inner products.    
  
\par 
 As for the second observation, recall that a turn $T(\bm{\hat{n}},\,\alpha)= \exp  
 \,[-i\alpha\, \bm{\hat{n}}\cdot \bm{\tau}]$ acts on the Poincar\'{e} sphere ${\cal  
 P}$ as rotation of amount $2\alpha$ about the directed axis $\bm{\hat{n}}$. This may  
 be viewed as continuous evolution for a time duration $\alpha$ under the {\em  
 constant} Hamiltonian $\bm{\hat{n}}\cdot \bm{\tau}$. Under this rotation or  
continuous evolution, states on ${\cal P}$ are driven on circles of constant  
 latitude about $\bm{\hat{n}}$. One of these orbits is a great circle. Evolution on  
 these orbits, or circles of constant latitude, is {\em not an in-phase evolution in  
general}.   The geodesic orbit is the only exception: A state on ${\cal P}$ located  
orthogonal to  $\bm{\hat{n}}$ evolves in such a way that successive states are in  
phase with one another. As    
an illustration, assume $\bm{\hat{n}} = (0,\,0,\,1)$ so that    
\bea \label{uz}  
T(\bm{\hat{n}},\,\alpha)_{\bm{\hat{n}}=(0,0,1)} \leftrightarrow \exp \,[-i\alpha\tau_3]  
= \left(\,\begin{array}{cc}\;\cos\alpha\; & \;- \sin\alpha\\\sin\alpha&\cos\alpha    
\end{array}\,\right).\nonumber\\  
\eea  
The distinguished great circle in this case is the equator and the relevant states are again the linearly polarized states considered under the first observation above.   
That the $2\times 2$ {\em real Jones matrix} in Eq.\,(\ref{uz}) drives these   
{\em real Jones vectors} in an in-phase manner is obvious.   
That our claims hold for a general $\bm{\hat{n}}$ and the associated great circle follows from unitary equivalence. This may be loosely paraphrased as follows. Under the unitary evolution driven by $T(\bm{\hat{n}},\,\alpha)$, states on the great circle orthogonal to $\bm{\hat{n}}$   
evolve, but not their phases. States on the other constant latitude circles   
evolve with a corresponding evolution of phases. The distinguished   
states at $\pm\bm{\hat{n}}\in {\cal P}$ do not evolve; their phases alone evolve   
by $ \pm\alpha$.   
\begin{figure}[ht]   
\includegraphics[width=6.0cm]{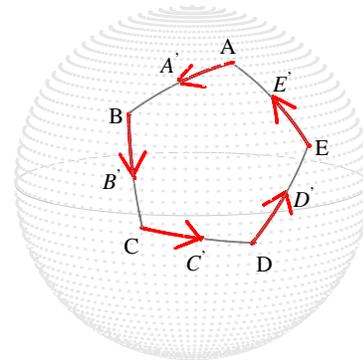}  
\caption{\footnotesize{Shown, in the case of a geodesic pentagon on ${\cal P}$, is the connection between Pancharatnam phase and Hamilton's turns. ${\rm A',\;B',\,C',\;D',\; E'}$ are the mid-points of the sides of the pentagon ABCDE. On the one hand turns ${\rm AA',\;BB',\;CC',\;DD',\;EE'}$ evolves or drives the initial state A  in an in-phase manner along the five sides of the pentagon, resulting in phase change   
(Pancharatnam phase) equaling half the area of ABCDE. on the other hand, these turns compose to a turn about A of turn length equaling half the area of ABCDE. }}  
\label{fig6}  
\end{figure}   
\par  
The connection between turns and Pancharatnam or geometric phase emerges quite simply when we   
combine these two observations.  Consider the geodesic pentagon ABCDE of states   
shown in Fig.\,\ref{fig6}. Let ${\rm A',\;B',\;C',\;D',\;E'}$ be the mid points of, respectively, AB, BC, CD, DE, and EA, and let ${\cal A}$ be the area of   
ABCDE. Starting with the state corresponding to the point A,   
${\rm turn~AA'}$ drives it in an in-phase manner along AB to B,       
${\rm turn~BB'}$ drives B in an in-phase manner along BC to C,       
${\rm turn~CC'}$ drives C to D along CD,       
${\rm turn~DD'}$ drives D to E, and finally        
${\rm turn~EE'}$ drives E  back to A along EA  in an in-phase manner.   
\par  
Now this closed-circuit evolution can be {\em viewed from two perspectives}.     
 Since the state A is taken along the pentagon ABCDE back to A {\em in   
an in-phase manner}, according to Pancharatnam the final state at A    
will differ from the initial one by a phase equal to ${\cal A}/2$. From the perspective of   
Hamilton and his turns, we have that ${\rm turn~AA'}$,   ${\rm turn~BB'}$,  ${\rm turn~CC'}$,    
 ${\rm turn~DD'}$, and  ${\rm turn~EE'}$ acting in that sequence   
have the combined effect of leaving  A invariant. That is, A is a fixed point of    
${\rm turn~AA'}+ {\rm turn~BB'}+ {\rm turn~CC'}+{\rm turn~DD'}+  {\rm turn~EE'}$.   
However, any turn leaving A invariant should be a turn about $\bm{\hat{n}}_A$, the unit   
vector specified by A\,$\in {\cal P}$. Since we know from the area   
formula (\ref{uxx}) adapted to the present case     
 that the turn length  of this composite turn is ${\cal A}/2$, we conclude     
\bea  
T(\bm{\hat{n}}_A,\,{\cal A}/2)= {\rm turn~AA'}+ {\rm turn~BB'}+ {\rm   
turn~CC'}&&\nonumber\\  
    +{\rm turn~DD'}+  {\rm turn~EE'},~~~~&&   
\eea  
and this completes the connection between Hamilton's turns and Pancharatnam's   
geometric phase.  
\par 
Our demonstration has been for the case of a pentagon, but it   
should be clear that the conclusion generalizes to $n$-sided polygons and, as a suitable limit, to any closed circuit on ${\cal P}$.  
  
\section{Elementary exercises in the use of turns}  
 
In this section we illustrate the use of turns in several simple  
situations involving single-qubit gates. The insight developed through these elementary exercises will prove to be of much value in our analysis   
to be taken up in the next section.  
\begin{figure}[ht]  
\includegraphics[width=5.5cm]{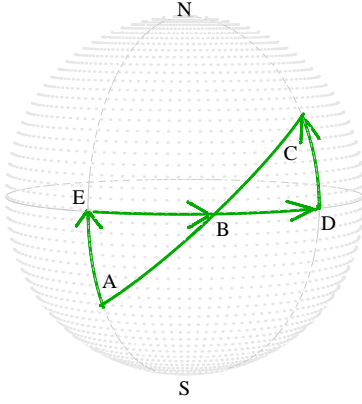}  
\caption{\footnotesize{The relationship between the Euler   
parameters $\xi,\eta,\zeta$ of an SU(2) gate   
$u(\xi,\,\eta,\,\zeta)$ and the positional  
spherical coordinates on ${\cal T}$   
of the associated turn. The positional coordinates of B,\,C,\,A are,   
respectively, $(\pi /2,\,\varphi_1),\,(\theta,\,\varphi_2),\,  
(\pi-\theta,\,2\varphi_1-\varphi_2)$.} }  
 \label{fig7}  
\end{figure}   
\par  
Clearly, the first requirement for the effective use of turns as a  
tool-kit for handling  SU(2) gates is an ability to translate freely  
between the positional coordinates of a turn on the sphere ${\cal T}$   
on the one hand and the Euler parameters (angles) of the associated SU(2) matrix  
 on the other. Given an SU(2) gate $u(\xi, \eta,   
\zeta)$, we slide the representative arc of its turn on its great circle so that the tail  
(or head) is on the equator. Let the spherical coordinates $(\theta,  
\varphi)$ of the tail and head of turn BC be, respectively, $(\pi/2,  
\varphi_{1})$, $(\theta, \varphi_{2})$, as in Fig.\,\ref{fig7}.   
It is clear that ${\rm turn~BD  = turn~EB}$ corresponds to optical rotator  
$R(2\varphi_{2}-2{\varphi}_{1})$, ${\rm turn~DC}$ to birefringent plate  
$C_{-{\pi}/{4}+ {\varphi_{2}}/{2}}(\pi -2 \theta)$, and ${\rm turn~AE}$  
to $C_{-{\pi}/{4}+\varphi_1 - {\varphi_{2}}/{2}}(\pi-2 \theta)$. Thus,  
we have the suggestive resolution of ${\rm turn~BC}$ into its   
"vertical" (birefringence) and "horizontal" (optical rotation) parts\,:  
\begin{eqnarray}  
u(\xi, \eta, \zeta) & \sim & {\rm turn}~BC\,\,=\,\,{\rm turn}~BD  
\,\,+\,\, {\rm turn}~DC \nonumber \\  
& = & C_{-{\pi}/{4}+{\varphi_2}/{2}}(\pi -2 \theta)\,R(2 \varphi_2  
-2 \varphi_1).\;\;\;\,  
\end{eqnarray}   
\par 
In place of this rotation followed by birefringence decomposition we could  
have equally well considered the birefringence followed by  rotation  
decomposition. Since ${\rm turn~AB} = {\rm turn~BC}$, the positional   
coordinates of A are $(\pi-\theta,\,2\varphi_1-\varphi_2)$. We have   
\begin{eqnarray}  
u(\xi, \eta, \zeta) &\sim& {\rm turn}~AB\,\,=\,\,{\rm turn}~AE  
\,\,+\,\, {\rm turn}~EB \nonumber \\  
& =& R(2 \varphi_2 -2 \varphi_1)\,  
C_{-{\pi}/{4}+\varphi_1 -{\varphi_2}/{2}}(\pi -2 \theta).~~  
\end{eqnarray}  
Comparing either of these decompositions with Eq.\,(\ref{uxy})   
 one readily deduces   
\begin{eqnarray}  
\xi&=&-\frac{\pi}{2} +\varphi_{2} , \nonumber \\  
\eta&=& \pi - 2 \theta, \nonumber  \\  
\zeta &=& \frac{\pi}{2} + \varphi_2 - 2\varphi_1,  
\end{eqnarray}  
where $\varphi_1,\varphi_2,\theta$ are the positional  
spherical coordinates of the turn  
associated with $u(\xi,\eta,\zeta)$. 
\par 
These are precisely the kind of relationships we were after, and it is  
 significant that these expressions which connect the positional spherical   
coordinates on ${\cal T}$ of a turn  to its Euler angles  are {\em linear}.  
  
\begin{figure}[ht]   
\includegraphics[width=7.0cm]{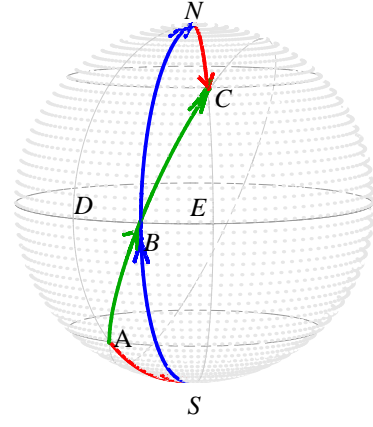}  
\caption{\footnotesize{The commutation relation between  
a QWP and a HWP,  yielding the rule for going from the Q-H configuration  
to the H-Q configuration and vice versa. }}  
\label{fig8}  
\end{figure}   
\par 
Our next exercise concerns the composition of a QWP and a HWP, as shown in  
Fig.\,\ref{fig8}. Points D, B, E are on the equator, A and C are on  
the circles of $45^{\circ}$  latitude, and N, S are the  
polar points. Let $\varphi_1$, $\varphi_2$, $\varphi_3$ be the azimuthal  
coordinates of the equatorial points D, B, E and assume DB = BE  
or, equivalently, $\varphi_3=2\varphi_2 -\varphi_1$. It is clear that ${\rm turn~AS}$   
represents $Q_{\pi/4 + \varphi_1 /2}$ while ${\rm turn~NC}$  
represents $Q_{{\pi}/{4}+{\varphi_{3}}/2}$. Further, ${\rm turn~SB}$ =  
${\rm turn~BN}$ represents $H_{-{\pi}/{4} +{\varphi_2}/{2}}$.  
\par 
Now consider the pair of spherical triangles ASB, CNB. The angle at  
N equals the angle at S, in view of the assumption $\varphi_3 -  
\varphi_2= \varphi_2 -\varphi_1$. Further,  AS = CN and SB = NB. Thus, these   
triangles are congruent, showing that ${\rm turn~AB = turn~BC}$. In other words,   
\bea  
{\rm turn~AS} + {\rm turn~SB} &=& {\rm turn~BN} + {\rm turn~NC}; \nonumber 
\eea 
that is, 
\bea 
H_{-\pi/4 +\varphi_2 /2}Q_{\pi/4 + \varphi_1 /2} &=& Q_{\pi/4 +   
\varphi_3 /2} H_{-\pi/4 + \varphi_{2} /2},\nonumber\\  
&&{\rm with}~ \varphi_1 + \varphi_3 = 2 \varphi_2.~~~  
\eea     
Denoting $\pi/4+\varphi_1/2 = \varphi$ and $-\pi/4 + \varphi_2/2 = \varphi^{\,'}$,  
the constraint $\varphi_1+\varphi_3 = 2\varphi_2$ is equivalent to $\pi/4 +  
\varphi_2/2 = 2\varphi^{\,'} - \varphi ~\;\text{mod}~\pi$, and so we have   
\begin{align}\label{ux1}  
H_{\varphi^{\,'}}\, Q_{\varphi} = Q_{2\varphi^{\,'}-\varphi}\, H_{\varphi^{\,'}}, ~  
\;\forall\; \varphi,\; \varphi^{\,'}.  
\end{align}  
This {\em "commutation relation"} shows that the H-Q configuration   
cannot have a capability not shared by the Q-H configuration.   
\par  
\begin{figure}[ht]   
\includegraphics[width=5.5cm]{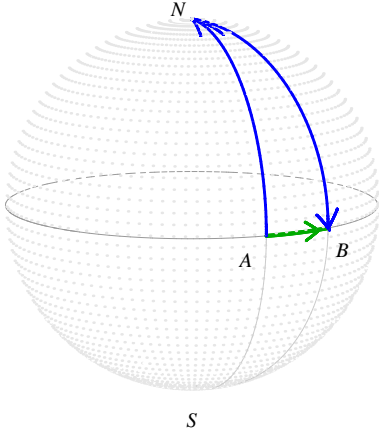}  
\caption{\footnotesize{Realization of variable optical rotator using a pair of HWPs.   
  This can also be viewed as depicting the special ability of a HWP to  
"absorb" an arbitrary   
 amount of optical rotation.} }  
\label{fig9}  
\end{figure}   
The composition of a pair of HWPs is our next exercise and this is  
depicted in Fig.\,\ref{fig9}. Let $\varphi_1, \, \varphi_2$ be the azimuthal  
coordinates of the equatorial points A,\,B. Then $ {\rm{turn}~AB} =  
R(2\varphi_2-2\varphi_1)$,  ${\rm turn~SA}={\rm turn~AN} = H_{-\pi/4+\varphi_1/2}$,  
and ${\rm turn~NB}={\rm turn~BS}=H_{\pi/4+\varphi_2/2}$. One readily reads  
out from Fig.\,\ref{fig9}  
\begin{align}  
\rm{turn}~AB&=\rm{turn}~AN + \rm{turn}~NB \nonumber\\  
\text{that is, }~~ R(2\varphi_2-2\varphi_1) &= H_{\pi/4+\varphi_2/2} \,  
H_{-\pi/4+\varphi_1/2}.  
\end{align}    
With $\pi/4+\varphi_2/2 \equiv \varphi$ and $-\pi/4 + \varphi_1/2 \equiv  
\varphi^{\,'}$ we have   
\begin{align}  
H_{\varphi}\,H_{\varphi^{\,'}} = R(2\pi + 4(\varphi-\varphi^{\,'})).  
\end{align}  
Recall from Eq.\,(\ref{ux5}) that it is $R(4\pi)$, and not $R(2\pi)$,  
that equals the SU(2) identity $\tau_0$.  
 [Indeed, $R(2\pi) = -\tau_0$.]  Noting that   
$H_{\pm \pi/2 +\varphi}$ is the inverse of $H_{\varphi}$, we may rewrite the   
last identity in the form   
\begin{align}\label{uxz}  
H_{\varphi}\,H_{\pm \pi/2+ \varphi^{\,'}} = R(\,4(\varphi-\varphi^{\,'})\,).  
\end{align}  
The preceding identity [Eq.\,(\ref{uxz})] shows  that a variable optical rotator can be simply realized with a pair of HWPs, the effective rotation or optical activity being linear  
in the relative orientation (of the fast axis) of the HWPs.   
\par  
There is another instructive and important manner in which   
Fig.\,\ref{fig9} can be read. Since $\rm{turn~BN}$ corresponds to $H_{-\pi/4+\varphi_2/2}$, the  fact that $\rm{turn~AN} = \rm{turn~AB}+\rm{turn~BN}$ reads $H_{-\pi/4+\varphi_1/2} =  
H_{-\pi/4+\varphi_2/2}\, R(2\varphi_2-2\varphi_1)$, or $H_{\varphi}\,R(\alpha) =  
H_{\varphi-\alpha/4}$. Similarly, the visual identity $\rm{turn~SA} + \rm{turn~AB} =  
\rm{turn~SB}$ reads $R(2\varphi_2-2\varphi_1) \,H_{-\pi/4 + \varphi_1/2} =  
H_{-\pi/4+\varphi_2/2}$. We have thus proved   
\begin{align}  
R(\alpha)\, H_{\varphi} = H_{\varphi+\alpha/4}, \,~~  H_{\varphi}\,R(\alpha) =  
H_{\varphi-\alpha/4}.   
\end{align}   
These two identities are equivalent to, and consistent with, one another in view of   
the defining  
property $R(\alpha)\, H_{\varphi} \,R(-\alpha) =   
\Phi(\alpha/2)\, H_{\varphi} \,\Phi(-\alpha/2) = H_{\varphi + \alpha/2}$ of  
$R(\alpha)$, and they exhibit {\em the special capability of a HWP to  
`absorb' optical rotation and still remain a HWP}. This absorption   
property, combined with the earlier noted property that a pair of HWPs is  
simply equivalent to an optical rotator, implies that three HWPs can  
fare no better than one HWP.   
\par 
Our last result shows that any number of HWPs cannot, by themselves,  
realize much portion of the manifold of SU(2) gates; indeed, an odd  
number of them is  no better than just one HWP, while an even number is simply   
equivalent to a (variable) optical rotator. In either case, not more than a one-parameter family of SU(2) gates gets realized.   
\begin{figure}[ht]  
\includegraphics[width=5.5cm]{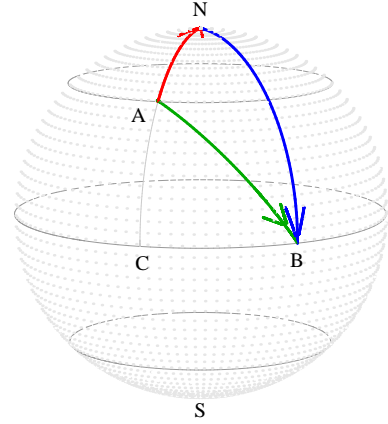}  
\caption{\footnotesize{The subset of unitary gates that can be realized with one QWP   
and one HWP. Shown also is the fact that addition of a second HWP   
can not in any manner enlarge   
this subset. }}  
\label{fig10}  
\end{figure}   
\par   
One is thus led to ask how much portion of the SU(2) manifold of  
unitary gates can be realized with two HWPs and one QWP. Let us begin  
with just one HWP and one QWP, as shown in  Fig.\,\ref{fig10}, where the point A  
lies on the $45^{\circ}$ latitude circle, and B and C on the equator. Let  
$\varphi_1$ be the azimuthal coordinate of A and C and $\varphi_2$  
 that of B. Then $\rm{turn~AN}$ corresponds to $Q_{-\pi/4+\varphi_{1}/2}$ and  
$\rm{turn~NB}$ to $H_{\pi/4+\varphi_2/2}$. As a consequence of the visual   
identity   
$\rm{turn~AB} = \rm{turn~AN}+\rm{turn~NB}$ we deduce that ${\rm turn~AB}$ corresponds to  
$H_{\pi/4+\varphi_2/2} Q_{-\pi/4+\varphi_1/2}$. It follows that the turns or  
SU(2) gates realizable with one HWP and one QWP are distinguished by  
the property that such turns extend from one of the $45^{\circ}$ latitude circles to   
the  equator or, equivalently, from the equator to a $45^{\circ}$ latitude circle.   
\par 
However, from the identity $\rm{turn~AB}=\rm{turn~AC} + \rm{turn~CB}$ we  
see that such a turn is equivalent to a QWP followed by an optical  
rotator. Since an optical rotator and a pair of HWPs are equivalent,  
we conclude that turns realizable by two HWPs and one QWP are  
certainly of the same form as $\rm{turn~AB}$, that is, extending between  
a $45^{\circ}$ latitude circle and the equatorial circle. Since such turns  
are fully parametrized by their azimuthal coordinates on these two  
circles, we conclude that two HWPs and one QWP can realize only a  
two-parameter family of SU(2) gates, and that this family is  
precisely the one realized using one QWP and just one HWP.   
\par 
It turns out that with two QWPs and one HWP one can realize the  
entire SU(2) manifold of unitary gates [that three QWPs by  
themselves will not suffice is readily seen from the fact that they cannot  
realize, for instance,  any SU(2) element whose turn has length $>3\pi/4$]. Before  
we turn to the next section for proof of this claim of three component realization  
of {\em all} SU(2) gates we examine, as our last exercise in this section, the manner  
in  
which a pair of QWPs transform a variable optical rotator into a variable birefringent  
plate.   
\begin{figure}[ht]   
\includegraphics[width=6.0cm]{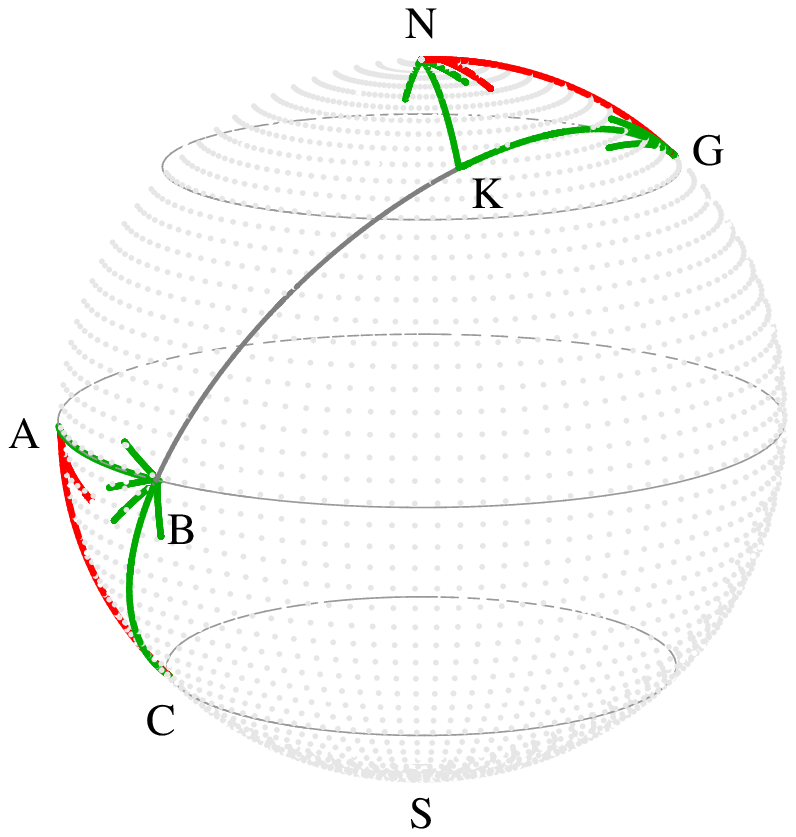}  
\caption{\footnotesize{Diagram showing that a pair of QWPs can convert variable optical rotation into variable birefringence.}}  
\label{fig11}  
\end{figure}   
\par  
The points $\rm{C,\, G }$ of  Fig.\,\ref{fig11} are on the $45^{\circ}$ latitude   
circles and points ${\rm A,\,B}$ are on the equator. Let ${\rm AB} = \eta/2$. Point ${\rm K}$ on the geodesic ${\rm CBG}$ is so chosen that ${\rm CB=KG}$. It is clear that  
$\rm{turn~GN}$ corresponds to QWP $Q_0$, $\rm{turn~CA}$  
to its inverse, $Q_{\pi/2}$, and $\rm{turn~AB}$ to $R(\eta)$.  
\par  
Let us now consider the pair of spherical triangles ${\rm CAB, \, GNK}$. We  
have ${\rm AC=NG}$, angle at ${\rm C}=$ angle at ${\rm G}$,  while ${\rm CB= KG}$ by  
construction. Thus, the triangles are congruent. This means, on the one  
hand, that angle at ${\rm N}=$ angle at ${\rm A=}~\pi/2$ so that $\rm{turn~KN}$ is a  
birefringent plate $C_{-\pi/4}(\cdot)$ and, on the other hand, that $\rm{KN=AB=}~\eta/2$, so  
that $\rm{turn~KN}$ corresponds to $C_{-\pi/4}(\eta)$. We have thus  
established   
\begin{align}  
\rm{turn}~KN&=\rm{turn}~KG+\rm{turn}~GN \nonumber\\  
&=\rm{turn}~CB + \rm{turn}~GN \nonumber \\  
&= \rm{turn}~CA+ \rm{turn}~AB + \rm{turn}~GN\,; \nonumber\\  
\text{that is, }~~~ C_{-\pi/4}(\eta) &= Q_0\, R(\eta)\, Q_{\pi/2}.  
\end{align}         
Conjugating by $\Phi(\pi/4)$ we have   
$C_0(\eta) = Q_{\pi/4}\, R(\eta)\, Q_{-\pi/4}$, which on conjugation by   
$\Phi(\varphi)$ yields   
\begin{align}  
C_{\varphi}(\eta) = Q_{\pi/4+\varphi}\, R(\eta)\, Q_{-\pi/4+\varphi}.  
\end{align}  
This relationship between  variable optical rotation and variable  
birefringence  is what we set out to demonstrate. However, to the extent that the  
variable optical rotator is not a preferred component, the fact remains that this may not be the most convenient way to realize variable birefringence.   
\section{Realization of all SU(2) gates Using Two QWPs and one HWP}  
Since HWP and QWP have each only one (rotational) degree of freedom, and since SU(2) is   
a 3-parameter manifold, it is clear that at least three components are required to realize even a small nontrivial (nonzero measure) part of this manifold. We have already noted that one QWP and two  HWPs cannot fare any better than a QWP plus a HWP. In the present section we prove, using  insights developed through the elementary exercises of the last section, that two QWPs and one HWP are sufficient to realize the full group manifold of SU(2) gates.  
\par 

\begin{figure}[ht]  
\includegraphics[width=6.0cm]{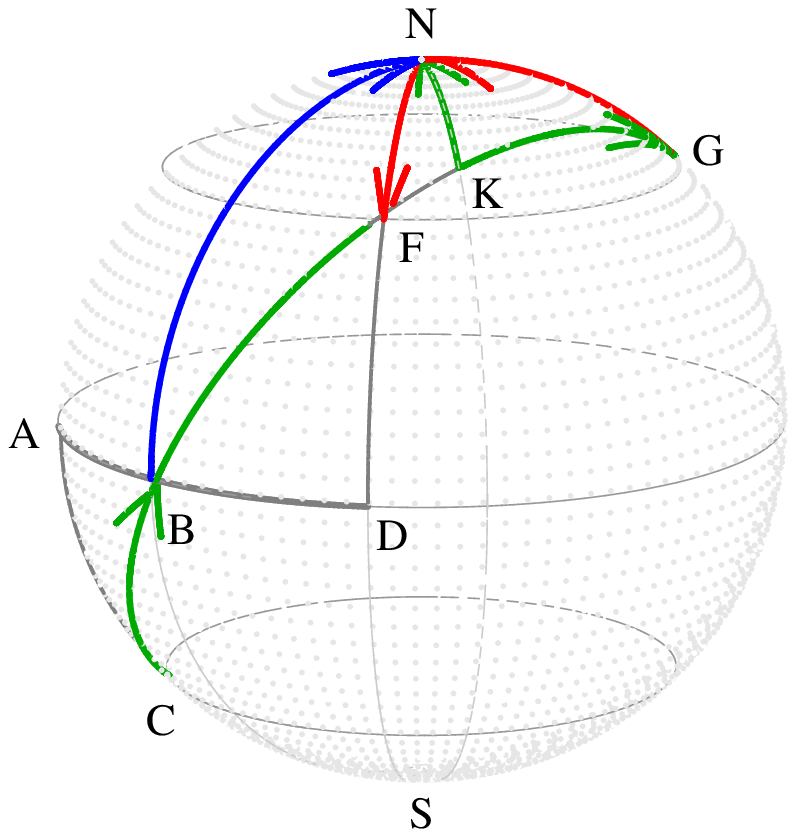}  
\includegraphics[width=6.25cm]{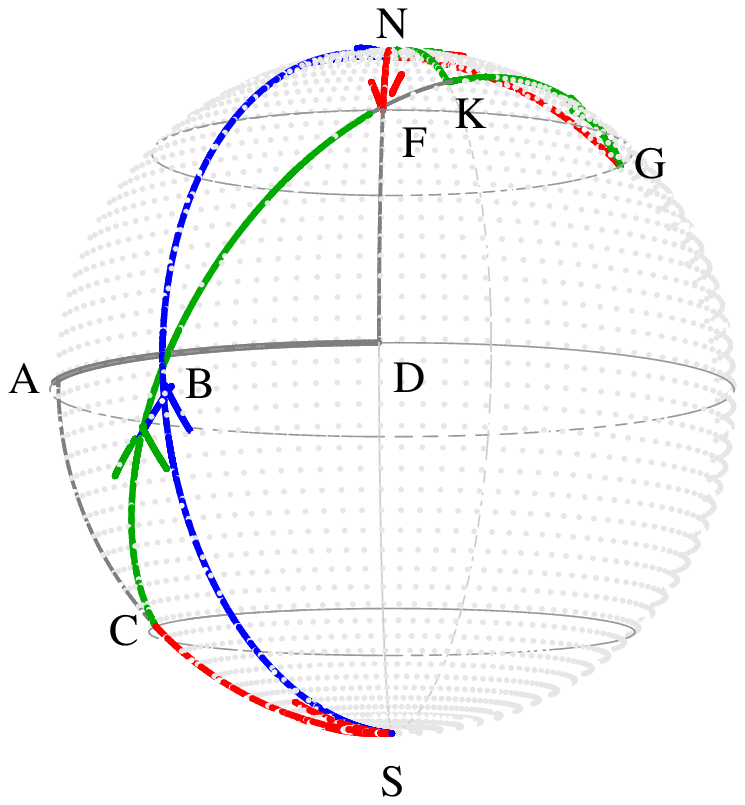}  
\caption{\footnotesize{Realization of all SU(2) gates using   
just two QWPs and one HWP. Two perspectives are presented for   
visual convenience. }}  
 \label{fig12}  
\end{figure}

The proof is straight-forward and relies entirely on Fig.\,\ref{fig12}. The points A, B, D are on the equator, N and S are the polar points, and C, G are on the circles of 45$^\circ$ latitude. The point F, the intersection of line (great circle arc) DN with line CBG, is {\em not assumed} to be on the circle of 45$^\circ$ latitude; the fact that F indeed lies on this   
circle will emerge by itself. The point K on the line CBG is so chosen that turn~CB = turn~KG. We {\em do not assume} that the angle GNK equals $\pi/2$. The fact that NK and NG   
are orthogonal at N will unfold on its own; turn~KN  will then correspond to a   
birefringent plate $C_{-\pi/4}(\eta^{\,'})$, where $\eta^{\,'}$ equals twice the arclength of   
KN. It is this turn corresponding to variable birefringence   
that will eventually become the focus of our attention.  
\par  
We assume ${\rm AB} = {\rm BD} =\eta/2$. Our first task is to prove that the  
point F  
lies  
on the 45$^\circ$ latitude circle. We begin by noting that turn~GN corresponds to  QWP $Q_0$  and so is also turn~CS.  
\par 
Consider  the pair of spherical triangles ABC, DBF. The angle at A equals the angle at  
D (both equal $\pi/2$). Both triangles have the same angle at B,  and AB = BD by   
 construction. Thus, the two triangles are congruent. It follows that  DF= AC =  
$\pi/4$,  
showing that F  lies indeed on the 45$^\circ$ latitude circle, and that turn~BF = turn~CB. Since F is on the 45$^\circ$ latitude circle, turn~NF corresponds to QWP $Q_{\eta/2}$.  Since AB = $\eta/2$, turn~BN = turn~SB corresponds to  HWP   
$H_{\pi/2+\eta/4} = H_{-\pi/2+\eta/4}$.  
\par 
Now consider the pair of spherical triangles ABC, KNG. The angle at C   
obviously equals the angle at G. Further, CB = KG by construction and AC = GN (= $\pi/4$), showing that the two triangles are congruent. As one consequence we see that the angle at N is   
$\pi/2$, showing that turn~NK is indeed the birefringent plate $C_{-\pi/4}(\,\cdot\,)$.   
As another consequence we have KN = AB = BD = $\eta/2$, showing that turn KN =   
$C_{-\pi/4}(\eta)$. But, in a visually obvious manner,    
\bea  
{\rm turn~KN} &=& {\rm turn~KG} + {\rm turn~GN}\nonumber\\   
&=& {\rm turn~BF} + {\rm turn~GN} \nonumber\\  
 &=& {\rm turn~BN} + {\rm turn~NF} + {\rm turn~GN}.\;\;  
\eea  
Since turn~KN = $C_{-\pi/4}(\eta)$,  and since the three turns on the right-hand side of the last equation equal, respectively,    
$H_{\pi/2+\eta/4}$,   $Q_{\eta/2}$,  and $Q_{0}$, we have proved  
\bea \label{ux2}  
C_{-\pi/4}(\eta)= Q_{0}\, Q_{\eta/2}\, H_{\pi/2+\eta/4},  
\eea  
{\em demonstrating the realizability of variable birefringence}.  
We see from Fig.\,\ref{fig12} that turn~KN could have also been developed in a somewhat different but equivalent manner:  
\bea  
{\rm turn~KN} &=& {\rm turn~KG} + {\rm turn~GN}\nonumber\\   
&=& {\rm turn~CB} + {\rm turn~GN} \nonumber\\  
 &=& {\rm turn~CS} + {\rm turn~SB} + {\rm turn~GN}.\;\;  
\eea  
It is clear that this corresponds to the multiplicative identity   
\bea  
C_{-\pi/4}(\eta)  
= Q_{0}\, H_{\pi/2+\eta/4}\,Q_0.  
\eea	  
Incidentally, the fact that these two realizations of  $C_{-\pi/4}(\eta)$   
 respectively in the Q-Q-H and Q-H-Q configurations are equivalent can   
also be verified using the H-Q commutation relation in Eq.\,(\ref{ux1}).   
\par  
Conjugating Eq.\,(\ref{ux2}) by $\Phi(\pi/4)$, which corresponds to rigidly rotating   
Fig.\,\ref{fig12} by $\pi/2$ about the polar axis,  we have  
\bea  
C_{0}(\eta)= Q_{\pi/4}\, Q_{\pi/4+\eta/2}\, H_{-\pi/4+\eta/4},   
\eea	  
a variable birefringent plate with the fast polarization   
along the $x_1$ axis. The reader will appreciate that Fig.\,\ref{fig12}   
was crafted  for $C_{-\pi/4}(\eta)$ rather than $C_{0}(\eta)$   
simply for visual clarity.  
\par  
Conjugating the last equation by $\Phi(\xi/2)$, which amounts to a further rigid   
rotation of Fig.\,\ref{fig12} by $\xi$ about the polar axis, one obtains  
\bea  
C_{\xi/2}(\eta)= Q_{\pi/4+\xi/2}\, Q_{\pi/4+\xi/2+\eta/2}\,   
H_{-\pi/4+\xi/2+\eta/4}.\;\;  
\eea	  
Right multiplying by the optical rotator $R(\xi+\zeta)$, and using the special   
ability of HWP to "absorb" rotation, namely $H_{(\cdot)}R(\alpha) = H_{(\cdot)   
-\alpha/4}$,  we have in view of  $u(\xi,\,\eta,\,\zeta)= C_{\xi/2}(\eta)R(\xi+\zeta)$   
  given in Eq.\,(\ref{uxy})   
\bea\label{ux6}  
u(\xi,\,\eta,\,\zeta) = Q_{\pi/4+\xi/2} \,Q_{\pi/4+\xi/2+\eta/2}\,   
H_{-\pi/4+ (\xi+\eta-\zeta)/4},\;\,\nonumber\\  
\eea	  
showing that all SU(2) gates can be realized using two QWPs and one HWP in the   
Q-Q-H configuration. Uniqueness of this realization as well as the fact that the required   
 (angular) positions of the plates are linear in the Euler angles 
should be appreciated.

That a similar assertion holds for the other two possible configurations, namely, Q-H-Q and   
H-Q-Q, follows immediately from the Q-H commutation relation   
$Q_\alpha H_\beta = H_\beta Q_{2\beta - \alpha}$ of Eq.\,(\ref{ux1}):  
\bea\label{ux3}  
u(\xi,\,\eta,\,\zeta) = Q_{\pi/4+\xi/2} \,  
H_{-\pi/4+ (\xi+\eta-\zeta)/4}\,  
Q_{\pi/4 - \zeta/2};\;\;\,  
\eea  
\bea\label{ux4}  
u(\xi,\,\eta,\,\zeta) =   
H_{-\pi/4+ (\xi+\eta-\zeta)/4}\,  
Q_{\pi/4+(\eta-\zeta)/2} \,  
Q_{\pi/4 - \zeta/2}. \nonumber\\  
\eea	  
We have thus completed a proof of the main result of this section,   
indeed this paper:   
  
\noindent  
{\bf Theorem}: All SU(2) gates $u(\xi,\,\eta,\,\zeta)$   
 can be realized with just two QWPs and one HWP equally well in any one of the three conceivable configurations Q-Q-H, Q-H-Q, or H-Q-Q, as  detailed, respectively, in  Eqs.\,(\ref{ux6}), (\ref{ux3}), and (\ref{ux4}). The realization is unique in each case, and in each configuration the angular positions of the plates are linear in the Euler angles $\xi,\,\eta,\,\zeta$ of the gate.  
 \begin{figure}[ht]  
\includegraphics[width=7.5cm]{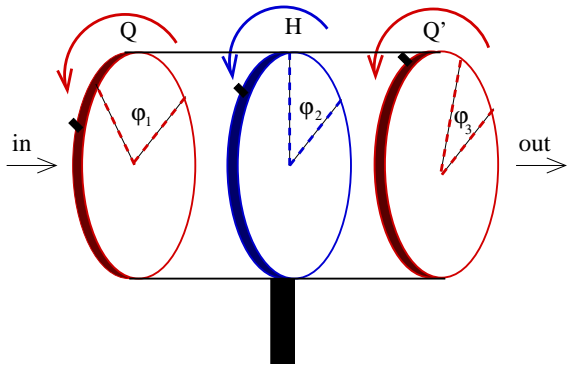}  
\caption{\footnotesize{The assembly of the universal gadget   
which realizes all single-qubit unitary gates. The three wave plates  
Q, H, Q' are coaxialy mounted, with a provision being made to assign and read   
the orientations of their fast axes any triplet of values.}}  
 \label{hturns13}  
\end{figure}   
\section{Universal single-qubit unitary gate}  
The theorem proved above  enables assembling of a universal   
single-qubit unitary gate as  described below. While the universal   
gate can be assembled in any one of the three configurations Q-Q-H, Q-H-Q, or H-Q-Q we choose the configuration Q-H-Q simply in order to be concrete.   
\par  
Let two QWPs Q, Q$^{\,'}$ and a HWP H be coaxially   
mounted in a cylindrical case, as shown in Fig.\,\ref{hturns13}. Assume that a provision  
is made to   
rotate each one of the three plates Q, H, Q$^{\,'}$ independently about the axis of the cylinder, and that provision is made to read out the angular coordinates   
$\varphi_1,\; \varphi_2,\; \varphi_3$ of their fast axes on their respective   
 (semicircular)  dials. We assume further that the assembly is so used that light passes through Q, H, Q$^{\,'}$ in that order.   
\par  
Clearly, the SU(2) matrix corresponding to the entire assembly is   
$Q^{\,'}_{\varphi_3}\,H_{\varphi_2}\,Q_{\varphi_1}$. Thus, to realize   
any unitary gate $u(\xi,\,\eta,\,\zeta)\in~$SU(2) we have to simply arrange these dial positions to read   
\bea  
\varphi_1 &=& \pi/4 -\zeta/2~~ {\rm mod}\;\pi,\nonumber\\  
\varphi_2 &=& -\pi/4 +(\xi+\eta-\zeta)/4~~ {\rm mod}\;\pi,\nonumber\\  
\varphi_3 &=& \pi/4 +\xi/2~~ {\rm mod}\;\pi,  
\eea  
as may be seen from Eq.\,(\ref{ux3}).  
 \par 
That the entire SU(2) manifold of unitary gates can be realized   
using a single assembly has its advantages. For instance, if the only   
imperfections of the wave plates from ideal ones are residual losses,   
the fact that the total loss of the assembly is the same for realization   
of every SU(2) gate, independent of the Euler angles 
$(\xi,\eta,\zeta)$  
of the gate, implies that it can be  
accounted for more easily.   
 Further, the fact that the dynamical phase through the system remains  
the same for all gates can prove to be of particular importance   
in interference considerations, particularly in the context of   
geometric phase experiments\,\cite{kimble1,mandel}.   \\

\section{Concluding Remarks}  
In this paper we have developed in considerable  
 detail the pictorial construction  
 of Hamilton into a tool-kit for handling problems of synthesis and analysis  
 in situations where the unitary group SU(2) plays a central role.  
 This group pervades nearly all areas of science,  
 either directly or through its cousin SO(3). Thus, this tool-kit should be of  
interest to a wider audience beyond quantum computation and quantum  
information. [In particular, the formalism and results presented here  
should be  of much interest to classical polarization optics.] 
 It is for this reason that we have developed Hamilton's construction  
itself  
in sufficient detail, taking care  to bring out its connection with geometric  
phase and the Bargmann invariants. It is in view of this detailed preparation  
that we believe the manipulations with turns presented in Sections~6, 7, and 8  
 will be found to be fully accessible to a broad spectrum of readership.
\par
As noted in the Introduction, the central result presented as a theorem  
towards the end of the paper is not new in itself.  
Our presentation is fashioned to  
work toward this result for two reasons. First, to demonstrate how this   
 geometric approach is suggestive and visually transparent compared to the algebraic  
approach.  Second, this result acts as a focal point in the sense that in working  
towards it most of the simple manipulations with turns are conveniently   
developed and demonstrated in stages.
\par
It is true that all the results developed here geometrically could be  
algebraically verified through matrix multiplication. However, it is in the  
geometric representation that synthesis results  {\em suggest} themselves in  
a vivid or visual manner. We may cite the H-Q commutation relation  
in Fig.\,8 and the "absorption" of optical rotation by a HWP  
(as also the realization of variable optical rotator using a pair of HWPs) in  
Fig.\,9 as effective illustrations of this advantage.  
\par
Finally, our presentation of turns in this paper is  
 in the context of the unitary group SU(2). The reader  
will easily realize that this geometric representation readily translates to  
the rotation group  SO(3) if a turn of length $\ell$ and the {\em reversed}   
turn of length $\pi-\ell$ are identified.  This amounts to identifying  
 the null turn with the turn of length $\pi$, which is clearly the  
same thing as identifying $U$ of SU(2) with $-\,U$ of SU(2). Thus,  
the turn length $\ell$ in the case of SO(3) gets restricted to the range  
$0\le\ell  \le \pi/2$.  

\vskip 0.2cm

\noindent
{\sf Acknowledgement}: The authors would like to thank Professor N. Mukunda 
for a critical reading of the manuscript.


\end{document}